\RequirePackage{fix-cm}
\documentclass[smallextended]{svjour3}       
\smartqed  
\usepackage[british]{babel}
\usepackage{url}
\usepackage[breaklinks=true,colorlinks]{hyperref}
\usepackage[hyphenbreaks]{breakurl}
\hypersetup{final}
\usepackage{graphicx}
\usepackage[utf8]{inputenc}
\usepackage{listings}
\usepackage{multicol}
\usepackage{multirow}
\usepackage{colortbl}
\usepackage{xcolor}
\usepackage{graphicx}
\usepackage[frozencache,cachedir=minted-cache]{minted}
\usepackage[figuresright]{rotating}
\usepackage[caption=false]{subfig}
\usepackage{tikz,pgf}
\let\svtikzpicture\tikzpicture
\def\tikzpicture{\noindent\svtikzpicture}

\begin{document}
\title{Demystifying Regular Expression Bugs
}
\subtitle{A comprehensive study on regular expression bug causes, fixes, and testing}


\author{Peipei Wang         \and
        Chris Brown \and 
        Jamie A. Jennings \and
        Kathryn T. Stolee
}


\institute{Peipei Wang, Chris Brown, Jamie A. Jennings, Kathryn T. Stolee \at
              Department of Computer Science, North Carolina State University, 
              Raleigh, NC, USA\\
              \email{pwang7@ncsu.edu, dcbrow10@ncsu.edu, jjennings@ncsu.edu, ktstolee@ncsu.edu}
}

\date{Received: date / Accepted: date}
\maketitle

\begin{abstract}
Regular expressions cause string-related bugs and open security vulnerabilities for DOS attacks. 
However, beyond ReDoS (Regular expression Denial of Service), little is known about the extent to which regular expression issues affect software development and how these issues are addressed in practice. We conduct an empirical study of 356 merged regex-related pull request bugs from Apache, Mozilla, Facebook, and Google GitHub repositories. We identify and classify the nature of the regular expression problems, the fixes, and the related changes in the test code. 

The most important findings in this paper are as follows: 
1) incorrect regular expression behavior is the dominant root cause of regular expression bugs (165/356, 46.3\%). 
The remaining root causes are incorrect API usage (9.3\%) and other code issues that require regular expression changes in the fix (29.5\%), 
2) fixing regular expression bugs is nontrivial as it takes more time and more lines of code to fix them compared to the general pull requests, 
3) most (51\%) of the regex-related pull requests do not contain test code changes. Certain regex bug types (e.g., compile error, performance issues, regex representation) are less likely to include test code changes than others, and 
4) the dominant type of test code changes in regex-related pull requests is test case addition (75\%).
The results of this study contribute to a broader understanding of the practical problems faced by developers when using, fixing, and testing regular expressions. 

\end{abstract}

\section{Introduction}
\label{intro}
Regular expression research in software engineering has explored performance issues~\cite{cody2017search,wang2019hyperscan}, comprehension~\cite{chapman2017exploring}, translation between languages~\cite{davis2019aren,davistesting}, and test coverage~\cite{peipei2018}. These efforts are motivated by the assumption that regular expressions are pervasive in systems. 
For example, through the lens of GitHub issues, a simple search for “regex OR regular expression” yields 227,474 results (and growing), with 25\% of those still being open. 
Regular expressions are reported to cause 37\% of string-related bugs~\cite{eghbalino},  
and yet, regular expressions are poorly tested~\cite{peipei2018}. 

This work aims to demystify regular expression bugs and to uncover the nature of the issues that relate to regular expressions, and in particular, the nature of the issues that developers end up addressing. This work further explores the role of test code in bug fixing for regex-related bugs.

As a lens into issues developers face, we explore merged pull requests (PRs) related to regular expressions (\emph{regex-related pull requests}). The assumption is that pull requests that are merged represent issues in code that developers find worthy of fixing. We target large open-source projects -- specifically Apache, Mozilla, Google, and Facebook -- that use the pull request model for code contributions. This allows us to study the problem, solution, and discussions about regular expressions in multiple programming languages. Prior work suggests that there are significant differences in some regex characteristics across programming languages~\cite{davistesting}, and our findings echo this: we likewise find differences in bug characteristics across languages. 

The study of bug descriptions and code changes in the regex-related pull requests leads to one of our main findings: a classification of regular expression bugs addressed by developers. For example, developers write regular expressions that are too constrained three-times as often as they write regular expressions that are too relaxed. This has implications for test case generation research, indicating the importance of generating strings that are outside the regular expression language. 
While most of the related work on incorrect regex usage focuses on isolated regular expression strings and overlooks the context where regexes are used, this regex classification points out the importance of  context and reveals that incorrectly using regex API also causes multiple regex-related problems. 

As regular expressions are under-tested~\cite{peipei2018}, we are interested in the impact of regular expression bugs on the testing effort. That is, are developers motivated to write tests if there is a bug in their regular expression? Thus, we explore the test code changes alongside the regex-related changes in PRs. 
From the test case changes, we can observe developers' behaviors on testing regular expressions, which gives hints on how to improve regular expression testing.

The contributions of this work include:
\begin{itemize}
    \item The first comprehensive empirical study on regular expression bugs in real-world open-source projects across multiple languages.
    \item Identification of root causes and manifestations of 356 regular expression bugs in 350 merged pull requests related to regular expressions. 
    \item Analysis of regular expression bug fix complexity and the connection between root causes and the changes in a fix.
    \item Ten common patterns in regular expression bug fixes.
    \item A quantitative and qualitative analysis of pull request test code changes, illustrating their relationship with various types of regular expression issues.
\end{itemize}

This paper extends prior work~\cite{wang2020empirical} by studying the test code changes in the regex-related pull requests. Besides the contributions listed above regarding test code, we provide test code comparisons with other datasets and benchmarks, explanations for why many regex-related pull requests do not have test code changes, and clues as to why it is hard to test regular expressions.

The rest of the paper is organized as follows. Section~\ref{sec:rqs} proposes our research questions which are answered in Section~\ref{sec:bug}, Section~\ref{sec:fix}, and Section~\ref{sec:rq3}, respectively. Section~\ref{sec:study} shows how we collect and analyze data. Section~\ref{sec:discuss} discusses our observations and the implications of this study. Section~\ref{sec:threats} presents threats to validity. Section~\ref{sec:related} presents the related work and  Section~\ref{sec:conclusion} concludes this paper.

\section{Research Questions}
\label{sec:rqs}
The goal of this study is to understand the regular expression bugs developers address in practice. We obtain our data via purposely selected GitHub pull requests and carefully analyze these pull requests to achieve this goal. Specifically, this study asks and answers the following questions:  
\begin{description}
\item[RQ1:] {\em What are the characteristics of the problems being addressed in regex-related PRs?}

We use an open card sort to categorize the root causes of the problems that pull requests deal with.  Three root causes emerge: 1) the regex itself; 2) regex API; and 3) other code. Within each type of root cause, we further characterize different manifestations of the addressed problem and provide more details about each manifestation (see Section~\ref{sec:bug}). 

\item[RQ2:] {\em What are the characteristics of the fixes applied to regex-related PRs?}

In analyzing the fixes in regex-related PRs, 
we measure fix complexity with four PR features proposed in prior work~\cite{gousios2014dataset}: 1)~minutes between PR opening and merging; 2) the number of commits in the PR; 3) the number of lines changed in the fixes; and 4) the number of files touched in the fixes. 
We then zoom in to study the four types of regex-related code changes: 1) regex edit; 2) regex addition; 3) regex removal; and 4) API changes. For each PR root cause and manifestation, we identify the dominant type of change. 
Finally, we identify ten common fix patterns to fix either a regex bug or a regex API bug (see Section~\ref{sec:fix}). 

\item[RQ3:] {\em What are the characteristics of the test code in regex-related PRs?}

Through analyzing the test code involved in regex-related PRs, we analyze whether a pull request has test code changes along with the regex-related code changes. If the PR does contain test code changes, using the granularity of test cases, we classify the changes as: 1) test case edit; 2) test case addition; or 3) test case removal. The information we collected from the test code splits the dataset into two groups: PRs not containing test code changes through which we understand why test code changes are not involved in the fixes, and PRs containing test code changes through which we get the distribution of change types (see Section~\ref{sec:rq3}). 
\end{description}

\section{Study}
This section describes the data collection process and analyses to address RQ1, RQ2, and RQ3. 

\label{sec:study}
\subsection{Dataset}
Our dataset is a sample of merged GitHub pull requests. We chose merged GitHub pull requests for two reasons: 1)~our study is oriented towards the existing solutions of regular expression issues. Compared to GitHub issues, merged pull requests provide us with both the problem description and a solution; 
and 2) merged pull requests indicate the priority of the regular expression issues and the feasibility to fix them, which are not always satisfied by GitHub issues since they may cover very general regular expression discussions or Q\&As and thus do not embody a direct solution. 

\subsubsection{Artifact Collection}
As we aim to focus on real resolutions to real bugs, we examined repositories from established organizations with relatively mature development processes and active projects. These repositories have many commits, contributors, and culture around pull request use. 
We targeted four large active GitHub organizations: Apache~\cite{apache}, Mozilla~\cite{mozilla}, Google~\cite{google}, and Facebook~\cite{facebook}. 
Using the GitHub GraphQL API~\cite{GitHubGraphQL}, we searched for merged pull requests\footnote{ While we avoid many perils of mining GitHub~\cite{kalliamvakou2014promises} through our selection of organizations and projects (i.e., Perils II, III, IV, V, and VI), evaluating only merged pull requests is Peril VIII and thus a threat to validity, as discussed in Section~\ref{sec:threats}.} with ``regular expression" or ``regex" in the title or description with the last update time before February 1st, 2019.
We selected only repositories that have  Java, JavaScript, or Python as the primary language, as these are the three most popular programming languages used on GitHub~\cite{githut}. 
This resulted in 664 merged pull requests from 195 GitHub repositories in the 4 organizations.

\subsubsection{Pruning}
We limited our focus to pull requests that are {\bf regex-related}. A PR is called {\em regex-related} only if there are changes to a regular expression or a regular expression API method. In regex-related PRs, there is at least one regular expression that is added, removed, or edited, or there is at least one modification to regex APIs. 
For example, Figure~\ref{fig:regexAddition} shows an example of the regex \mintinline{javascript}{/\d+/} being added on line 4. We manually inspected the 664 merged PRs and identified 350 of them (52.7\%) as regex-related PRs. 

\begin{figure}[tb]
    \caption{Example of Regex Addition from a pull request in JavaScript~\href{https://github.com/mozilla/zamboni/pull/442}{(mozilla/zamboni\#442)}}
    \label{fig:regexAddition}
\begin{minted}[framerule=.5\textwidth,fontsize=\small,linenos]{javascript}
 gettext(format('Changes in {0} {1}',
         this.app.trans[this.app.guid],
-        this.app.version.substring(0,1)))));
+        /\d+/.exec(this.app.version)))));
\end{minted}
\vspace{-15pt}
\end{figure}

\subsubsection{Final Dataset Description}
The final dataset of 350 regex-related PRs comes from 135 GitHub repositories. Of these, 86 are from Apache repositories, 162 are from Mozilla repositories, 66 are from Facebook repositories, and 36 are from Google repositories. 
When analyzing regex-related code changes, we considered the overall code differences before and after the PR, hence avoiding issues from reworked commits (Peril VII~\cite{kalliamvakou2014promises}). 
Because a pull request can handle multiple independent regular expression problems, six PRs are split, creating a final dataset with 356 bugs addressed by pull requests, or \emph{pull request bugs}. Our final data are available~\cite{Dataset}. 



\begin{sidewaysfigure}[!tpb]
\subfloat[PR~\href{https://github.com/mozilla/feedthefox/pull/43}{mozilla/feedthefox\#43}]{
\label{fig:a}
\centering
\begin{tikzpicture}
    \draw (0, 0) node{\includegraphics[width=0.45\textwidth]{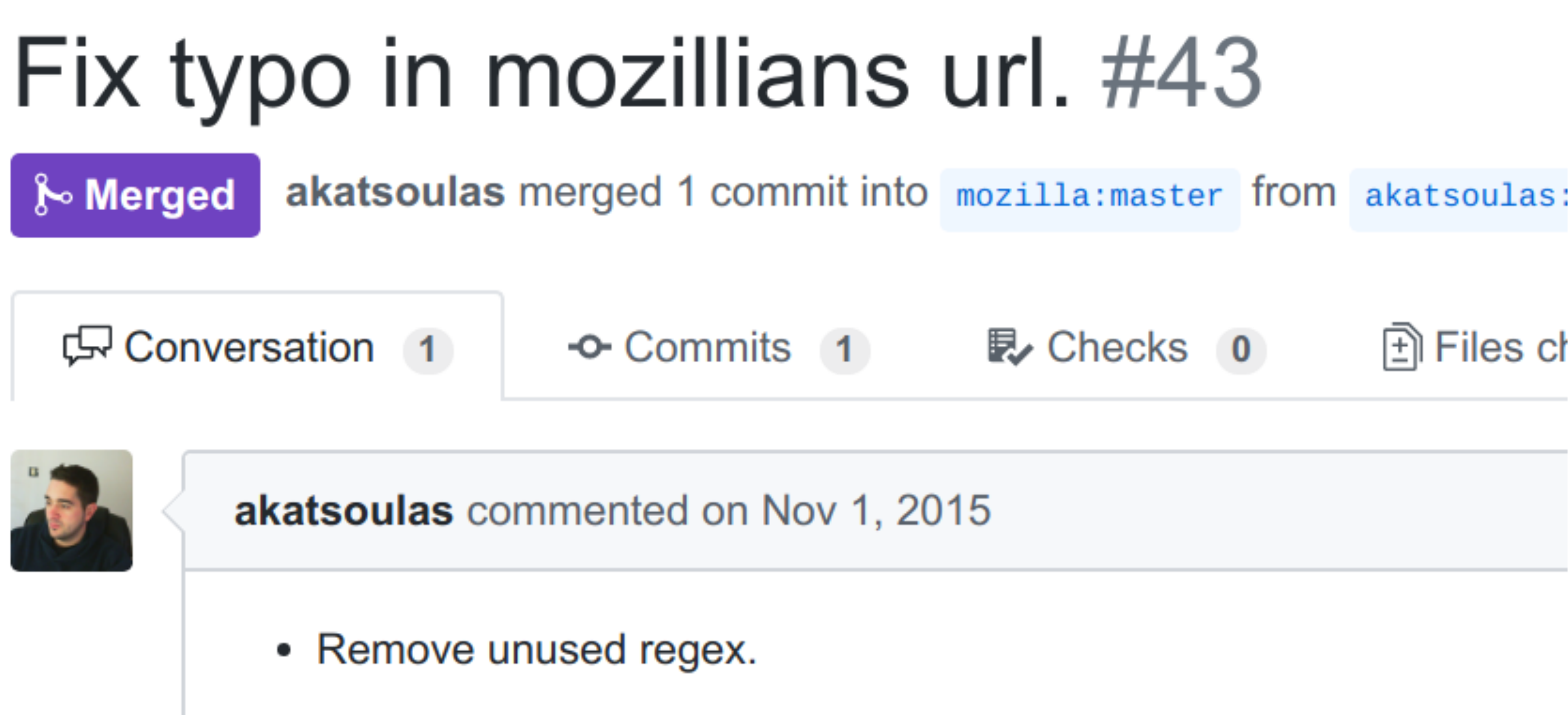}};
    \draw [line width=0.25mm, red] (-2.90,-1.35) rectangle (0.5,-1.85);
    \draw [line width=0.25mm, red](-4.35,1.25) rectangle (1.65,1.85);
\end{tikzpicture}
}
\quad
\subfloat[The GitHub issue for PR~\href{https://github.com/mozilla/addons-server/pull/10352}{mozilla/addons-server\#10352}]{
\label{fig:b}
\centering
\begin{tikzpicture}
    \draw (0, 0) node{\includegraphics[width=0.52\textwidth]{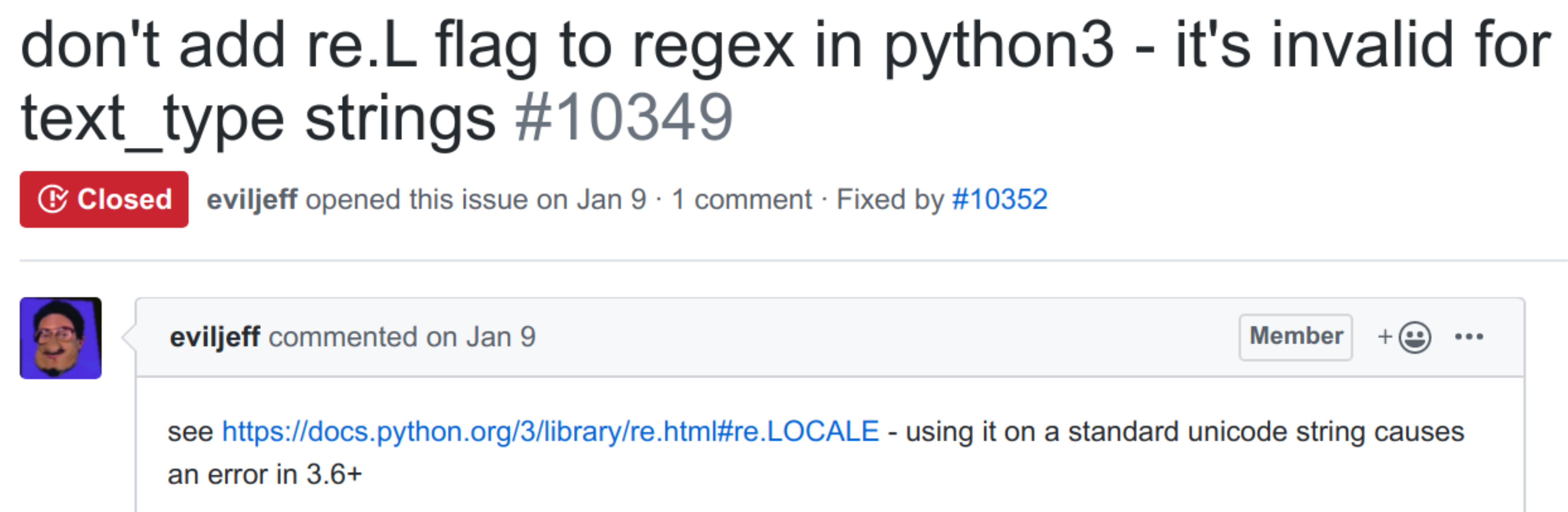}};
    \draw [line width=0.25mm, red](-4.05,-0.9) rectangle (4.45,-1.6);
\end{tikzpicture}
}
\vfill
\subfloat[The JIRA issue for PR~\href{https://github.com/apache/ambari/pull/760}{apache/ambari\#760}]{
\label{fig:c}
\centering
\begin{tikzpicture}
    \draw (0, 0) node{\includegraphics[width=\textwidth]{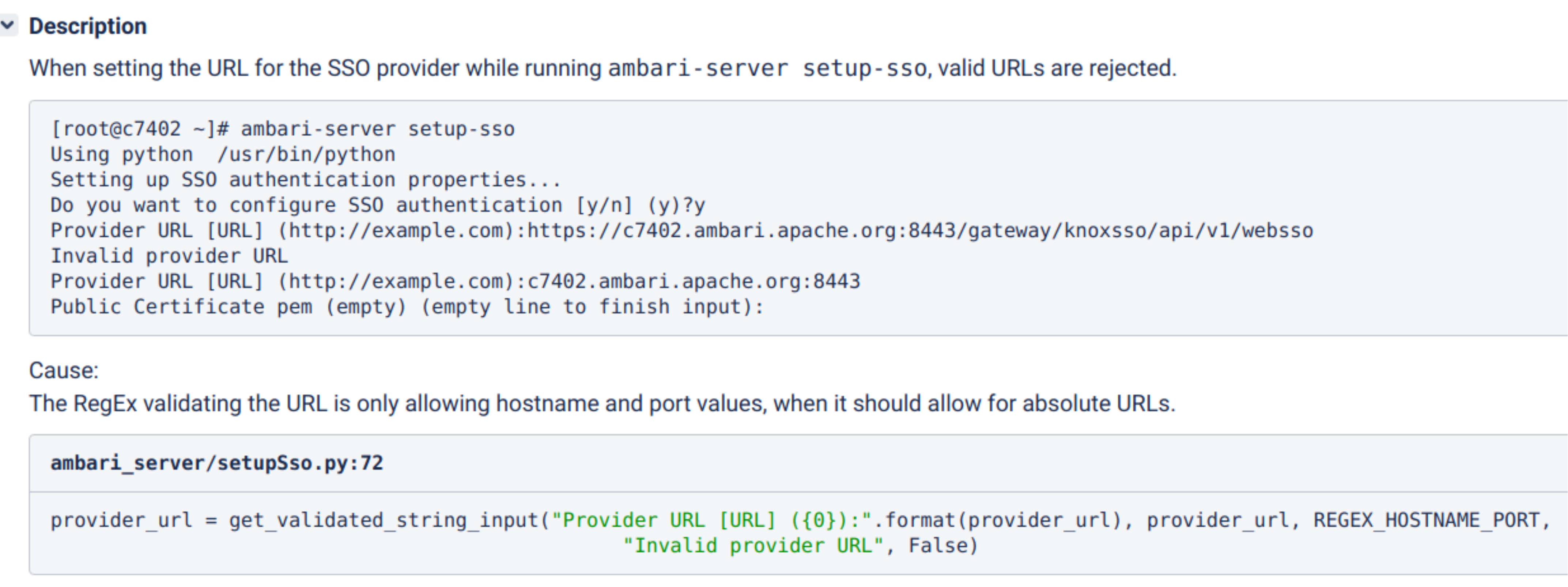}};
    \draw [line width=0.25mm, red](-9.30,-1.1) rectangle (4.80,-1.6);
    \draw [line width=0.25mm, red](-9.30,3.0) rectangle (4.80,2.5);
\end{tikzpicture}
}
\caption{Examples to illustrate identifying problems addressed in pull requests.}
\label{fig:problems}
\end{sidewaysfigure}

\subsection{RQ1 Analysis: Bug Characteristics}

With the 356 pull request bugs, two authors performed an open card sort with two raters. 
The dataset is categorized in two dimensions, \textit{root cause} and \textit{manifestation}, based on the pull request description, comments, linked GitHub issues, or linked bug reports from other systems (e.g., JIRA, Bugzilla). 
{\color{blue}Figure~\ref{fig:problems} shows three illustrative example PRs. }
In Figure~\ref{fig:a}, PR~\href{https://github.com/mozilla/feedthefox/pull/43}{mozilla/feedthefox\#43} addresses two problems. One is a typo of a variable shown in the title of this pull request, the other problem is an unused regex shown in the description of this pull request. We ignore the typo problem because the fix to the typo does not involve any regex or API changes. In the analysis of this PR, the fix is to remove the regular expression, and the problem it addresses is {\em unused regex} which is a type of {\em regex code smell}. 
Figure~\ref{fig:b} shows a PR where the addressed problem is described in a separate GitHub issue, which identifies an error caused by an incorrect flag in the {\em regex API} with the manifestation of {\em exception handling}~\footnote{The specific error message is ``ValueError: cannot use LOCALE flag with a str pattern". Since Python version 3.6, re.LOCALE can be used only with bytes patterns.}. 
Figure~\ref{fig:c} shows the JIRA bug report related to PR~\href{https://github.com/apache/ambari/pull/760}{apache/ambari\#760}. Per the highlighted text, the problem being addressed in this PR is {\em incorrect regex behavior} because valid URLs are rejected and the scope of the regular expression needs to be expanded. 

After card sorting was completed, eight manifestations of three root causes of regex-related bugs were identified. Four of the eight manifestations are further broken into categories and sub-categories according to the common characteristics shared by the bugs. The hierarchy of the 356 pull request bugs is presented in Table~\ref{tab:hierarchy}. 

\subsection{RQ2 Analysis: Fix Characteristics}
To answer RQ2, we explored regex fix characteristics compared to general software bugs, the nature of the changes in the fixes, and identify common fix patterns.

\subsubsection{Complexity of Regex-related PR Fixes}
\label{sec:analysis:rq2:complexity}
To understand if regex-related bugs are similar in complexity to other software bugs, we compared our regex-related PRs (\textit{regexPRs}) with a public dataset of PRs from GitHub projects that use PRs in their development cycle~\cite{gousios2014dataset}  (\textit{allPRs}). 
We selected four features from the prior work that represent the complexity of a fix or the complexity of reviewing a PR. To measure the complexity of reviewing the PR, we calculated the number of minutes from PR initialization to merging (\textit{mergetime\_minutes}). To measure the complexity of fixes in PRs, we chose the number of commits  (\textit{num\_commits}), the number of modified lines of code (\textit{code\_churn}), and the number of files changed  (\textit{files\_changed}). 
Note that {\em code\_churn} is a combined feature which is the sum of two originally proposed features, {\em src\_churn}, the number of lines changed in source code, and {\em test\_churn}, the number of lines changed in test code. This is because regular expressions are not only in source code but also in testing frameworks and configuration files, which makes it hard to distinguish the code of fixing a regex bug from the testing code.

The metrics for bug fix complexity in our dataset (\textit{regexPRs}) are obtained through the PyGithub~\cite{pygithub} library, which provides APIs to retrieve GitHub resources. 
The \textit{allPRs} dataset~\cite{gousios2014dataset} contains over 350,000 PRs; as a matter of fairness, we filtered out the unmerged pull requests and retained 300,600 merged ones for analysis. 
We used the Mann-Whitney-Wilcoxon Test~\cite{wilcoxon} to investigate whether our dataset, \textit{regexPRs}, and the \textit{allPRs} dataset have the same distribution. These comparison results are presented in Table~\ref{tab:pr_features}.
%

\subsubsection{Changes to Regexes in PRs}
We take into consideration four types of regex-related changes: 1) regular expression addition ($R_{add}$), 2) regular expression edit ($R_{edit}$), 3) regular expression removal ($R_{rm}$), and 4) regular expression API changes ($R_{API}$). 

Before counting the number of regex-related changes, we first identified regular expressions used in the code. 
Because the regular expression is often represented as a string or a sequence of characters, we treated each \emph{quoted} regex string as a normal string until we found it was parsed with regular expression syntax and a regular expression instance or object was created consequently. Strings wrapped by regular expression \emph{delimiters} are straightforward and treated as regular expressions. For example, slash \verb!/! in JavaScript is a regex delimiter. Hence \verb!/\d+/! in Figure~\ref{fig:regexAddition} is identified as a regex. 

A regular expression addition ($R_{add}$) is counted when the PR shows a new regular expression string. 
In the code snippet shown in Figure~\ref{fig:regexAddition}, there is no regex string prior to the PR whereas line 4 introduces regular expression \verb!/\d+/!. 

A regular expression edit ($R_{edit}$) is a content change to the regular expression string directly or indirectly used in regex API methods. These are the type of regular expression changes studied in prior work on regular expression evolution~\cite{peipei2019saner}. 

Similar to regex addition, a regular expression removal ($R_{rm}$) is counted when the code before a PR contains more regexes than after the PR. A pull request could directly remove a regex object (e.g., \href{https://github.com/mozilla/feedthefox/pull/43}{mozilla/feedthefox\#43}) or replace the regex and the code where it is used with other types of code (e.g., \href{https://github.com/google/graphicsfuzz/pull/167}{google/graphicsfuzz\#167}). 

 \begin{figure}[tb]
 \caption{Example of Regex API Changes from a pull request in Java~\href{https://github.com/google/ExoPlayer/pull/3185}{(google/ExoPlayer\#3185)}}
\begin{minted}[fontsize=\small,linenos]{java}
   currentLine = subripData.readLine();
-  Matcher matcher = SUBRIP_TIMING_LINE.matcher(currentLine);
-  if (matcher.matches()) {
+  Matcher matcher = currentLine == null ? null : 
        SUBRIP_TIMING_LINE.matcher(currentLine);
+  if (matcher != null && matcher.matches()) {
\end{minted}
\label{fig:regexAPI}
 \end{figure}
 
A regular expression API change ($R_{API}$) encapsulates changes to the APIs being used statically and dynamically. This includes modifying the method itself on a certain call site and reducing the execution frequency of that call site.
For modifying the API method, we counted only when the regex object shows up both before and after the PR. Therefore, API methods introduced with $R_{add}$ or removed with $R_{rm}$ are excluded. Take Figure~\ref{fig:regexAddition} as an example. In this example, the method \mintinline{javascript}{exec} is added as the side-effect of adding the regex \verb!/\d+/! and thus \mintinline{javascript}{exec} is not accounted as $R_{API}$. 
The modification to the method itself could be on its method name or arguments. 
If the modified argument is in the position for the regex string, it is not counted as an $R_{API}$  but as an $R_{edit}$. 
API changes also concern how the API methods are executed in run-time. For example, constructing regular expression objects statically rather than on-the-fly. The PR in Figure~\ref{fig:regexAPI} adds two checks of {\tt null} object, one for the argument passed into \mintinline{java}{Pattern.matcher} and the other for the instance invoking \mintinline{java}{Matcher.matches}. Hence, it is counted to have two regular expression API changes.
Another way of reducing call site execution frequency is to add guards (e.g., if-else statements) on the path of executing regular expression matching (e.g., \href{https://github.com/mozilla/treeherder/pull/61}{mozilla/treeherder\#61}).

\subsubsection{Recurring Patterns for Fixing Regular Expression Bugs}
To find the common fix patterns, we manually examined the code changes in pull requests caused by either regex or API. 
Since we are more interested in fixing regular expression bugs, the regex-related PRs caused by other code are out of scope for common fix patterns of regex bugs.
Each regex-related change is regarded as a different pattern, and similar changes are grouped together. We chose ten recurring patterns to represent the fix strategies for common regular expression problems (See Table~\ref{tab:patternsRegex}).


\begin{sidewaystable}[!tpb]
\begin{center}
\caption{The hierarchy for the 356 pull request bugs including root causes, manifestation, categories, and sub-categories.}
\label{tab:hierarchy}
\begin{tabular}{|c|l|l|l|r|r|r|} 
\hline
\multicolumn{1}{|l|}{Root Cause}                                      & Manifestation                       & \multicolumn{2}{l|}{Category~ (Sub-Category)}              & \multicolumn{1}{l|}{\begin{tabular}[c]{@{}l@{}}Count (\%) in\\(sub)Category \end{tabular}} & \multicolumn{1}{c|}{\begin{tabular}[c]{@{}c@{}}Count (\%) in \\Manifestation \end{tabular}} & \multicolumn{1}{c|}{\begin{tabular}[c]{@{}c@{}}Count (\%) in \\Root Cause \end{tabular}}  \\ 
\hline
\multirow{11}{*}{Regex}                                               & \multirow{5}{*}{Incorrect Behavior} & \multicolumn{2}{l|}{Rejecting valid strings}               & 102 (61.8\%)                                                                                & \multirow{5}{*}{165 (75.7\%)}                                                                & \multirow{11}{*}{218 (61.2\%)}                                                             \\ 
\cline{3-5}
                                                                      &                                     & \multicolumn{2}{l|}{Accepting invalid strings}             & 36 (21.8\%)                                                                                 &                                                                                             &                                                                                           \\ 
\cline{3-5}
                                                                      &                                     & \multicolumn{2}{l|}{Rejecting valid and accepting invalid} & 17 (10.3\%)                                                                                 &                                                                                             &                                                                                           \\ 
\cline{3-5}
                                                                      &                                     & \multicolumn{2}{l|}{Incorrect extraction}                  & 9 (5.5\%)                                                                                   &                                                                                             &                                                                                           \\ 
\cline{3-5}
                                                                      &                                     & \multicolumn{2}{l|}{Unknown}                               & 1 (0.6\%)                                                                                   &                                                                                             &                                                                                           \\ 
\cline{2-6}
                                                                      & \multicolumn{4}{l|}{Compile Error}                                                                                                                                                            & 8 (3.6\%)                                                                                    &                                                                                           \\ 
\cline{2-6}
                                                                      & \multirow{5}{*}{Bad Smells}         & \multirow{2}{*}{Design Smells} & Unnecessary regex         & 11 (24.4\%)                                                                                 & \multirow{5}{*}{45 (20.6\%)}                                                                 &                                                                                           \\ 
\cline{4-5}
                                                                      &                                     &                                & Other                     & 6 (13.3\%)                                                                                  &                                                                                             &                                                                                           \\ 
\cline{3-5}
                                                                      &                                     & \multirow{3}{*}{Code Smells}   & Performance issues        & 10 (22.2\%)                                                                                 &                                                                                             &                                                                                           \\ 
\cline{4-5}
                                                                      &                                     &                                & Regex representation      & 10 (22.2\%)                                                                                 &                                                                                             &                                                                                           \\ 
\cline{4-5}
                                                                      &                                     &                                & Unused/duplicated regex   & 8 (17.8\%)                                                                                  &                                                                                             &                                                                                           \\ 
\hline
\multirow{6}{*}{\begin{tabular}[c]{@{}c@{}}Regex\\API \end{tabular}}  & \multicolumn{4}{l|}{Incorrect Computation}                                                                                                                                                    & 6 (22.2\%)                                                                                   & \multirow{6}{*}{33 (9.3\%)}                                                                \\ 
\cline{2-6}
                                                                      & \multirow{5}{*}{Bad Smells}         & Design Smells                  & Alternative regex API     & 2 (7.4\%)                                                                                   & \multirow{5}{*}{27 (81.8\%)}                                                                 &                                                                                           \\ 
\cline{3-5}
                                                                      &                                     & \multirow{4}{*}{Code Smells}   & Unnecessary computation   & 9 (33.3\%)                                                                                  &                                                                                             &                                                                                           \\ 
\cline{4-5}
                                                                      &                                     &                                & Exception handling        & 8 (29.6\%)                                                                                  &                                                                                             &                                                                                           \\ 
\cline{4-5}
                                                                      &                                     &                                & Deprecated APIs           & 5 (18.5\%)                                                                                  &                                                                                             &                                                                                           \\ 
\cline{4-5}
                                                                      &                                     &                                & Performance/Security      & 3 (11.1\%)                                                                                  &                                                                                             &                                                                                           \\ 
\hline
\multirow{5}{*}{\begin{tabular}[c]{@{}c@{}}Other\\Code \end{tabular}} & \multirow{3}{*}{New Feature}        & \multicolumn{2}{l|}{Data processing}                       & 22 (37.3\%)                                                                                 & \multirow{3}{*}{59 (56.2\%)}                                                                 & \multirow{5}{*}{105 (29.5\%)}                                                              \\ 
\cline{3-5}
                                                                      &                                     & \multicolumn{2}{l|}{Regex-like implementation}             & 19 (32.2\%)                                                                                 &                                                                                             &                                                                                           \\ 
\cline{3-5}
                                                                      &                                     & \multicolumn{2}{l|}{Regex configuration entry}             & 18 (30.5\%)                                                                                 &                                                                                             &                                                                                           \\ 
\cline{2-6}
                                                                      & \multicolumn{4}{l|}{Bad Smells}                                                                                                                                                               & 19 (18.1\%)                                                                                  &                                                                                           \\ 
\cline{2-6}
                                                                      & \multicolumn{4}{l|}{Other Failures}                                                                                                                                                           & 27 (25.7\%)                                                                                  &                                                                                           \\ 
\hline
\multicolumn{6}{|c|}{Total}                                                                                                                                                                                                                                                                                                                                         & 356 (100\%)                                                                                \\
\hline
\end{tabular}
\end{center}
\end{sidewaystable}


\subsection{RQ3 Analysis: Test Code Characteristics}
\label{sec:test}
To answer RQ3, we identified the files used for test code and then manually counted the test cases being changed in the PR. While rare, if a PR contained more than one regex-related bug (six PRs in our dataset), we mapped the test code to the appropriate regular expression. 

\subsubsection{Identifying Test Code Files} 
Test code files are usually located under ``{\bf test}" (or ``{\bf tests}", ``{\bf \_\_test\_\_}``) and ``{\bf spec}" (or ``{\bf specs}") directories. Following certain naming conventions, the test file names could contain the word ``{\bf test}” or ``{\bf spec}" as prefix or suffix. Therefore, we regarded any files included in the PR which satisfy either of the two conditions as a candidate test code file. 

We exclude test code files that do not have an impact on testing the regex-related code logic. For example, the only change in file {\it HadoopFsHelperTest.java} is added comments in PR~\href{https://github.com/apache/incubator-gobblin/pull/63}{apache/incubator-gobblin\#63}. Therefore, this file is not included when counting test code changes of this PR. Similar changes not considered in test code files include code formatting, variable renaming, and moving code location. 

Note that a lack of test code changes in a PR does not mean test cases for the involved code do not exist. It just means that test code was not modified, added, or removed in the same PR.  

\subsubsection{Identifying Test Case Changes}
Once test files are identified, we can investigate the test code changes in more detail. 
We chose to classify changes at the test case level of granularity because it is used more often in test code studies~\cite{pinto2012understanding,beller2017oops,kochhar2013adoption} than test files and test classes~\cite{zaidman2008mining}.

The process for test case identification depended on the language and testing framework used. In the Python unit testing framework \mintinline{python}{unittest}~\footnote{\url{https://docs.python.org/3/library/unittest.html}} or \mintinline{python}{pytest}~\footnote{\url{https://docs.pytest.org/en/stable/}}, test method names start with the letters `{\bf test}'. In the Java Junit framework~\footnote{\url{https://junit.org/junit5/}}, test methods are annotated with the \mintinline{Java}{@Test} tag. In earlier Junit version 3, the unit test class is a subclass of \mintinline[]{java}{junit.framework.TestCase} with test methods prefixed with \mintinline{Java}{test}. For behavior-driven testing frameworks(e.g., \mintinline{javascript}{mocha}~\footnote{\url{https://mochajs.org/}}, \mintinline{javascript}{jasmine}~\footnote{\url{https://jasmine.github.io/}}), the \mintinline{javascript}{it} methods are regarded as the test case methods. Note that benchmark methods for testing regex-related code performance are also regarded as test cases. For the benchmark test files, each benchmark method is treated as a test case. 

To identify test case changes, we looked for new test cases added into the test code files ($T_{add}$), test cases with modifications ($T_{edit}$), and test cases removed from the test files ($T_{rm}$). 

If the inputs for tested methods are not in the test method body, changes in test inputs are considered to be test case changes as well. The inputs may be located in the same test file as the test case or located in separate files. 
Taking PR~\href{https://github.com/mozilla/treeherder/pull/181}{mozilla/treeherder\#181} as an example, file \textit{test\_tinderbox\_print\_parser.py} contains method \mintinline[breakanywhere]{python}{test_tinderbox_parser_output} where the test input source is \mintinline[breakanywhere]{python}{TINDERBOX_TEST_CASES}. Since the latter is added and it contains eight elements, $T_{add} = 8$ for this file with zero test case edits or removal. 
PR~\href{https://github.com/mozilla/treeherder/pull/334}{mozilla/treeherder\#334} is another example where test case changes are not directly reflected in the test code file but in the input source files of the test code. 
The modified two JSON files in this PR are the expected results of test case methods \mintinline{python}{test_mochitest_fail} and \mintinline{python}{test_jetpack_fail} on file {\it tests/log\_parser/test\_job\_artifact\_builder.py}. 
Although neither of the test method code changes, the expected results of those two files have changed. Therefore, this PR has $T_{edit} = 2$. 

Besides the changes to test case methods and test case input sources, we evaluated the impacts of test fixtures on test cases as well. PR~\href{https://github.com/apache/ambari-metrics/pull/8}{apache/ambari-metrics\#8} shows the modified test fixture \mintinline{java}{setUp()} contributes to the test case edits of \mintinline{java}{testReporterStartStop()} and \mintinline{java}{testMetricsExclusionPolicy()}. 

Similarly, we excluded code refactoring from the test code changes and also excluded test case changes that were made solely to pass the tests~\cite{pinto2012understanding}. 
In PR~\href{https://github.com/mozilla/fxa-auth-server/pull/1743}{mozilla/fxa-auth-server\#1743}, file \textit{sms.js} has a change from \mintinline{javascript}{/^US$/} to \mintinline{javascript}{[ `US' ]} because in the configuration the format of \mintinline{javascript}{regions} has changed from a \em{RegExp} to \em{Array}. 

\section{RQ1 Results: Bug Categories}
\label{sec:bug}
As is done in prior work on categorizing software bugs, we identified the \emph{root cause} and \emph{manifestation} of the bugs~\cite{zhang2018empirical,lu2008learning,di2017comprehensive,selakovic2016performance,tan2014bug}. The root cause is the location in the source code wherein the problem lies.
The manifestation is the impact of the bug on the code.

Among the 356 pull request bugs related to regular expressions, three root causes emerged: the \textbf{regex} itself (218, 61.2\%), the \textbf{regex api} used (33, 9.3\%), and \textbf{other code} (105, 29.5\%), as shown in the \emph{Root Cause} and \emph{Count (\%) in Root Cause} columns of Table~\ref{tab:hierarchy}. 
When the root cause is the \emph{regex}, the regex itself caused an issue; examples include incorrect behavior, a compile error, or a code smell. When the  \emph{regex api} is the root cause, this means the API was deprecated, the wrong flags were used, the API call is unprotected from exceptions, 
or another issue related to the use of the API is present.  When the root cause is \emph{other code}, the regex-related changes are identified but the fault or root cause lies elsewhere in the code (i.e., the regex or API was modified in a fix, but are not the root cause of the issue). 

Each root cause is divided by the manifestation of the bug, which describes how the bug was observed (\emph{Manifestation} and \emph{Count (\%) in Manifestation} columns of Table~\ref{tab:hierarchy}). 
For example, 45 PRs have \emph{Regex} as the root cause and manifest as a \emph{Bad Smell}, representing 20.6\% of the \emph{regex} root cause. Categories and sub-categories are used to further subdivide the manifestations (\emph{Category (Sub-Category)} and \emph{Count (\%) in Category} columns in Table~\ref{tab:hierarchy}). For example, 11 PRs have an \emph{Unnecessary Regex}, representing 24.4\% of the \emph{Bad Smells} for  the \emph{Regex} root cause.

Note that the manifestation of \emph{Bad Smells} appears for each of the root causes. This is because the PRs will frequently identify a better way to accomplish a behaviorally equivalent task, making the manifestation a bad smell rather than a fault. These bad smells, in aggregate, account for 91 (25.6\%) of the regex-related PR bugs.  Next, we describe each root cause category. 



\subsection{Bugs Caused by Regexes Themselves}
\label{sec:bug_regex}

When the regex is an issue (218 PR bugs), we observed three manifestations: \emph{incorrect  behavior}, \emph{compile error}, and \emph{bad smells}.

\subsubsection{Regex: Incorrect Behavior}
\textit{Incorrect Behavior} is the dominant manifestation for bugs with the regex as the root cause (75.7\%, 165/218). 
Table~\ref{tab:hierarchy} shows the four categories of this manifestation: rejecting valid string, accepting invalid strings, rejecting valid and accepting invalid strings, and incorrect extraction. 
Rejecting valid strings represents 61.8\% of the incorrect behavior bugs. 
This reinforces the observation that developers prefer to compose a conservative regex to an overly liberal one~\cite{michael2019regexes} and tend to expand the scope of regular expressions as software evolves~\cite{peipei2019saner}.


Two primary factors seem to contribute to incorrect regex behavior. 
One factor is incorrect regex escaping, including not escaping characters and incorrectly escaping characters such as backslash (\verb!\!) and forward slash (\verb!/!). 
The other is changing requirements. When the inputs change and the regex is not updated, the regex behavior may become obsolete (e.g., PR~\href{https://github.com/apache/cordova-ios/pull/376}{apache/cordova-ios\#376}). 
Other less common problems are related to case sensitivity, Unicode compatibility, misuses of quantifier greediness, and lack of anchors. 


\subsubsection{Regex: Compile Error}
Eight pull requests fix regex \textit{compile errors}. While the project code is compiled without errors, there could exist uncaught invalid regular expressions until runtime. For example, \href{https://github.com/apache/nutch/pull/234}{apache/nutch/\#234} reports a compile error caused by \mintinline{java}{File.separator} on Windows-based systems. Since \verb!\! is used for escaping other characters, this PR reports an uncaught \textit{PatternSyntaxException}. 


\subsubsection{Regex: Bad Smells}
\label{sec:regex:smell}
The regex \textit{bad smells} we observed can be divided into two categories, as shown in Table~\ref{tab:hierarchy}: \textit{design smells}, such as whether to use regex solution or not, which data to use for validation, and what the matching data and non-matching data look like; and \textit{code smells} referring to smells with the regex itself. 
Overall, 17 out of the regex bad smells are design smells and the other 28 are code smells. 

Most design smells were sub-categorized as \textit{unnecessary regex} (11/17). 
These PRs indicate that simpler solutions exist and a regex is not needed. 
For example, using a regex for string replacement is not necessary if the replaced string is a simple string literal in a fixed location (e.g., \href{https://github.com/mozilla/Snappy-Symbolication-Server/pull/23}{mozilla/Snappy-Symbolication-Server\#23}). 

The \textit{code smells} are roughly evenly distributed among three sub-categories. 
 {\em Performance issues} means the execution of regex could be optimized for speed or memory consumption. 
For example, when the purpose of a regex is not to extract substrings from the data input, defined capturing groups in the regex is unnecessary since the captured values are saved in memory but not used in later code (e.g.,\href{https://github.com/apache/struts/pull/156}{apache/struts\#156}). 
 Two of the performance issues are about regular expression complexity (i.e., ReDoS~\cite{davis2018impact} vulnerability\footnote{Since ReDoS cares about the time complexity of running the regular expression, we regard it as a performance issue.}).
\textit{Regex representation} means the regular expression string fails to satisfy certain unspecified requirements, such as using the raw string to describe regular expression in Python and following the eslint rule of ``No-regex-spaces"~\footnote{https://eslint.org/docs/2.0.0/rules/no-regex-spaces}.
Six of the ten regex representation code smells can be detected by lint tools in Python and JavaScript. 
The other four PRs fix one issue of escape characters in regex strings and three issues of regex readability. 
Unlike the incorrect behavior, the escape characters in this sub-category do not cause a behavioral issue. 
\textit{Unused/duplicated regex} refers to regexes in code that are no longer needed (7/8) or that are duplicated (1/8), with the former being more common. \\

\noindent\textbf{Summary:} Most incorrect regular expression behavior occurs when the regular expression is too conservative and needs to accept more strings. Compile errors occur in eight of the PRs, representing 2.2\% of all regex-related PRs we studied; considering the severity of compile errors in terms of disrupting the program execution, this is worth noting. Among design smells and code smells, 11 PRs identify the root cause as unnecessary regular expressions. 

\subsection{Bugs Caused by Regex APIs}
\label{sec:api}
Even with the correct regex, choosing the right API function is important, as is placing the API call in an appropriate location in the code. 
Bugs caused by regex APIs (33 PRs, 9.3\%) refer to the incorrect regex API usage manifesting as either \textit{incorrect computation} (6, 1.7\%) or \textit{bad smells} (27, 7.6\%). 


\subsubsection{Regex API: Incorrect Computation}
Six PRs were submitted because the API being used in the program produced incorrect results.  
For example, for a particular regular expression in \href{https://github.com/facebook/jest/pull/3001}{(facebook/jest\#3001)},  \mintinline{javascript}{RegExp.test(content)} has some unexpected behavior if it runs over the same string twice. This is because, in its context, the global matching flag \texttt{`g'} was used so the second call to this method starts matching from the position saved in the first call. This is a unique feature in JavaScript \emph{stateful} regex methods (i.e., \mintinline{javascript}{RegExp.test} and \mintinline{javascript}{RegExp.exec}). 
Besides the stateful methods, other incorrect API usage leading to incorrect computation includes passing arguments into the wrong method, failing to process multi-line inputs, and enforcing matching from the beginning or to the end of an input string. 


\subsubsection{Regex API: Bad Smells}
\label{sec:regexapi:smells}
We found 27 PR bugs that stem from \textit{bad smells} in using regex APIs. 
Table~\ref{tab:hierarchy} shows the breakdown of the regex API bad smells. 
Two design smells are \textit{alternative regex API} problems, such as deciding which regex library should be chosen to use (e.g.\href{https://github.com/facebook/prepack/pull/645}{facebook/prepack\#645}). 
The other 25 (92.6\%) are categorized as code smells.  

\textit{Unnecessary computation} was the root cause of nine PRs. In all cases, the issue is that the regex API is executed too many times and can be reduced. For example, on the code path where most of the jobs are a success, the regex parser for error messages should not be used unless the message indicates a job failure (e.g., \href{https://github.com/mozilla/treeherder/pull/61}{mozilla/treeherder\#61}). 
This is considered a regex API issue because it pertains to how the API is used in the code. It is a code smell because the code is behaving properly except for the performance. The frequency of this sub-category has implications for the impact regex API performance has on applications. 


\textit{Exception handling} refers to uncaught exceptions or errors in running regex methods. These represent issues with the regex APIs because developers did not account for the possible unexpected behaviors from executing a regex API. Examples include invalid regex syntax when the regex to compile is not hard-coded and unknown to the API method until runtime, invalid regex API method arguments (e.g., null values, unsupported regex flags), and invalid method returns (e.g., null values or incorrect return types). 

{\em Deprecated APIs} means an obsolete regex library is being used or there were changes in the new version of a regex library. 
For example, the old regex library \texttt{org.apache.oro} is replaced with \texttt{java.util.regex} (\href{https://github.com/apache/nutch/pull/390}{apache/nutch\#390})
because \texttt{org.apache.oro} has been retired since 2010 and users are encouraged to use Java regex library instead~\footnote{\url{https://jakarta.apache.org/oro/}}. 
Similarly, when flags argument is no longer supported~\footnote{\url{https://developer.mozilla.org/en-US/docs/Web/JavaScript/Reference/Global_Objects/String/replace}} in JavaScript regex APIs, \mintinline{javascript}{input.replace('<', '&lt;', 'g')} has to be changed into \mintinline{javascript}{replace(/</g, '&lt;')} (\href{https://github.com/mozilla/bugherder/pull/26}{mozilla/bugherder\#26}). 

\textit{Performance/Security} refers to a change in the API method due to performance or security concerns. For example, in JavaScript, developers found \mintinline{javascript}{regexp.test} to be more suitable than \mintinline{javascript}{str.match} because the former only returns a boolean value while the latter returns the matched results, which could create a leak of information to the external environment (\href{https://github.com/mozilla/hubs/pull/457}{mozilla/hubs\#457}).\\

\noindent\textbf{Summary:} Understanding the regex API is as important as understanding the regex itself. PR bugs result from choosing the wrong API~(6), using deprecated or updated APIs~(5), or improper exception handling~(8). Additional PRs reduce the number of calls to the regex API in the interest of performance~(9). 


\subsection{Bugs Caused by Other Code}
In these pull request bugs, regexes and their APIs are involved but are not the root causes of the bugs; the root cause is other code (105 PRs, 29.5\%). 
Regexes may be changed in these pull requests, but the regex is part of the solution, not part of the problem. 
For example, to solve a filename comparison failure \mintinline{javascript}{filename === 'jest.d.ts'} where the filename could be an absolute file path, a solution of regex matching is used to take the place (\href{https://github.com/facebook/react/pull/6804}{facebook/react6804}). 

The manifestations of the regex-related PRs caused by code other than the regex or the regex APIs are categorized according to how regex-related changes are involved in the solution. 
A PR is categorized as a {\em new feature} if it implements new functionality or improves existing features (59 PRs). 
Note that we also regard feature improvement as a new feature. 
A PR is categorized as a \textit{bad smell} if the regular expression is employed to refactor the source code and to remove the smells (19 PRs). 
A PR is categorized as {\em other failures} if it reports any other failure (27 PRs). 


\subsubsection{Other Code: New Feature}
Regular expressions are often involved in the introduction of new features. For example, to prevent malicious injection into logs, a regex is added to sanitize log messages (\href{https://github.com/apache/accumulo/pull/628}{apache/accumulo\#628}), which means the root cause is un-sanitized log messages, and sanitizing them is a new feature. 
Table~\ref{tab:hierarchy} shows category the breakdown of the 59 PRs for new features. 


\textit{Data processing}, which accounts for 22 PRs,  means the regular expression is added to process a specific type of data (e.g., \href{https://github.com/apache/drill/pull/452}{mozilla/bugbug\#65}). 
\textit{Regex configuration entry}, which accounts for 18 PRs, means the regex is user-provided so as to build regex-supported features satisfying different user needs (e.g., \href{https://github.com/apache/openwhisk-utilities/pull/16}{apache/openwhisk-utilities\#16}). 
\textit{Regex-like implementation} adds new functionality for performing regular expression execution. It requires both a regex and an input string but provides some unique features. 
For example, 
a data query engine added query methods (e.g., regexp\_matches) so that it can perform regex-like string searching in SQL queries (\href{https://github.com/apache/drill/pull/452}{apache/drill\#452}). 

\subsubsection{Other Code: Bad Smells}
When the root cause is a \textit{bad smell}, the solution is a refactoring; the regex or its API is involved with the refactoring. 
For example, a switch statement of over 85 cases can be refactored into less than 20 cases through the use of regexes (\href{https://github.com/apache/incubator-pinot/pull/2894}{apache/incubator-pinot\#2894}). 

\subsubsection{Other Code: Other Failures}
Regular expressions can also be added when the existing solution in the code does not work. 
For example, a regex solution can be used as a fix when the code of identifying browser type fails to identify a newer version of the browser (\href{https://github.com/mozilla/pdf.js/pull/7800}{mozilla/pdf.js\#7800}).\\

\noindent\textbf{Summary:} 
Regexes are involved in PRs even when the regex or its APIs are not the root cause. 

\begin{sidewaystable}[!htp]
\centering
\caption{Comparing selected features of regex-related PRs (\textit{regexPRs}) to merged PRs (\textit{allPRs}) from prior work~\cite{gousios2014dataset}.}
\label{tab:pr_features}
\begin{tabular}{|c|l|l|r|r|r|r|r|c|} 
\hline
Feature                             & \multicolumn{1}{c|}{Meaning}                                                                        & Dataset            & \multicolumn{1}{c|}{5\%} & \multicolumn{1}{c|}{mean} & \multicolumn{1}{c|}{median} & \multicolumn{1}{c|}{95\%} & \multicolumn{1}{c|}{skewness} & p-value       \\ 
\hline
\multirow{2}{*}{mergetime\_minutes} & \multirow{2}{*}{\begin{tabular}[c]{@{}l@{}}Minutes from PR \\initialization to merge~\end{tabular}} & \textit{allPRs}    & 0.00                     & 10,529.07                 & 405.00                      & 43,685.05                 & 10.99                         & -             \\ 
\cline{3-9}
                                    &                                                                                                     & \textit{regexPRs}  & 11.93                    & 10,212.00                 & 1,307.46                    & 47,589.74                 & 6.73                          & 8.139e-13***  \\ 
\hline
\multirow{2}{*}{num\_commits}       & \multirow{2}{*}{\begin{tabular}[c]{@{}l@{}}Number of commits\\in the PR\end{tabular}}               & \textit{allPRs}    & 1.00                     & 3.94                      & 1.00                        & 11.00                     & 16.75                         & -             \\ 
\cline{3-9}
                                    &                                                                                                     & \textit{regexPRs}  & 1.00                     & 2.67                      & 1.00                        & 8.00                      & 7.97                          & 0.3635        \\ 
\hline
\multirow{2}{*}{code\_churn}        & \multirow{2}{*}{\begin{tabular}[c]{@{}l@{}}Changed lines of \\code in the PR\end{tabular}}         & \textit{allPRs}    & 0.00                     & 324.15                    & 15.00                       & 1,047.00                  & 32.44                         & -             \\ 
\cline{3-9}
                                    &                                                                                                     & \textit{regexPRs}  & 2.00                     & 615.72                    & 27.00                       & 786.15                    & 18.01                         & 1.075e-08***  \\ 
\hline
\multirow{2}{*}{files\_changed}     & \multirow{2}{*}{\begin{tabular}[c]{@{}l@{}}number of files\\changed in the PR\end{tabular}}         & \textit{allPRs}    & 1.00                     & 11.84                     & 2.00                        & 30.00                     & 93.62                         & -             \\ 
\cline{3-9}
                                    &                                                                                                     & \textit{regexPRs}  & 1.00                     & 6.78                      & 2.00                        & 23.65                     & 8.20                          & 0.3068        \\
\hline
\end{tabular}
\newline 
*** p-value < 0.001 when comparing \textit{regexPRs} and \textit{allPRs} for that feature using the Mann-Whitney-Wilcoxon test.\\

\bigskip\bigskip  

\centering
\caption{Distribution of the four types of regex-related changes over different root causes and manifestations. A (B) means A PRs have B occurrences of the change, in total.
$\bullet$ indicates the dominant type of regex-related changes in the corresponding manifestation (or category) in each row. }
\label{tab:FixesOverall}
\begin{tabular}{|c|l|l|r|r|r|r|r|} 
\hline
\multicolumn{1}{|l|}{Root Cause} & \multicolumn{2}{l|}{Manifestation (Category)} & \#PR                     & \multicolumn{1}{c|}{ $R_{add}$ } & \multicolumn{1}{c|}{ $R_{edit}$ } & \multicolumn{1}{c|}{ $R_{rm}$ } & \multicolumn{1}{c|}{ $R_{API}$ }  \\ 
\hline
\multirow{5}{*}{Regex}           & \multicolumn{2}{l|}{Incorrect Behavior}       & 165                      & 22 (40)                          & 139 (236)$\bullet$                & 26 (48)                         & 12 (13)                           \\ 
\cline{2-8}
                                 & \multicolumn{2}{l|}{Compile Error}            & 8                        & 0 (0)                            & 7 (10)$\bullet$                   & 1 (3)                           & 3 (3)                             \\ 
\cline{2-8}
                                 & \multirow{2}{*}{Bad Smells} & Design Smells   & 17                       & 4 (5)                            & 4 (9)                             & 12 (63)$\bullet$                & 3 (4)                             \\ 
\cline{3-8}
                                 &                             & Code Smells     & 28                       & 3 (3)                            & 20 (49)$\bullet$                  & 8 (10)                          & 3 (5)                             \\ 
\cline{2-8}
                                 & \multicolumn{2}{c|}{Sum}                      & 218                      & 29 (48)                          & 170 (304)                         & 47 (124)                        & 21 (25)                           \\ 
\hline
\multirow{4}{*}{Regex API}       & \multicolumn{2}{l|}{Incorrect Computation}    & 6                        & 1 (1)                            & 1 (1)                             & 0 (0)                           & 6 (9)$\bullet$                    \\ 
\cline{2-8}
                                 & \multirow{2}{*}{Bad Smells} & Design Smells   & 2                        & 0 (0)                            & 0 (0)                             & 0 (0)                           & 2 (2)$\bullet$                    \\ 
\cline{3-8}
                                 &                             & Code Smells     & 25                       & 2 (8)                            & 3 (10)                            & 1 (25)                          & 23 (381)$\bullet$                 \\ 
\cline{2-8}
                                 & \multicolumn{2}{c|}{Sum}                      & 33                       & 3 (9)                            & 4 (11)                            & 1 (25)                          & 31 (392)                          \\ 
\hline
\multirow{4}{*}{Other Code}      & \multicolumn{2}{l|}{New Feature}              & 59                       & 53 (110)$\bullet$                & 3 (4)                             & 0 (0)                           & 4 (4)                             \\ 
\cline{2-8}
                                 & \multicolumn{2}{l|}{Other Failures}           & 27                       & 23 (44)$\bullet$                 & 6 (7)                             & 2 (4)                           & 3 (6)                             \\ 
\cline{2-8}
                                 & \multicolumn{2}{l|}{Bad Smells}               & 19                       & 11 (19)$\bullet$                 & 5 (21)                            & 5 (20)                          & 0 (0)                             \\ 
\cline{2-8}
                                 & \multicolumn{2}{c|}{Sum}                      & 105                      & 87 (173)                         & 14 (32)                           & 7 (24)                          & 7 (10)                            \\ 
\hline
\multicolumn{3}{|c|}{Total}                                                      & \multicolumn{1}{l|}{356} & 119 (230)                        & 188 (347)                         & 55 (173)                        & 59 (427)                          \\
\hline
\end{tabular}
\end{sidewaystable}

\section{RQ2 Results: Bug Fix Characteristics in Regex-related PRs}
\label{sec:fix}
While RQ1 describes the regex-related PR bugs, RQ2 describes the associated fixes. 
We approach this from three perspectives: 1) the complexity of the fix, compared to general PRs; 2) the types of changes to the code; and 3) frequently recurring bug fix patterns. 

\subsection{Complexity of Regex-related PR Fixes}

We hypothesize that regex-related PRs differ from most other PRs. We evaluate this hypothesis by comparing characteristics of regex-related PRs to PRs from a public dataset of representative PRs from GitHub projects that use PRs in their development cycle~\cite{gousios2014dataset}. 
Table~\ref{tab:pr_features} shows the pull request feature distributions for our dataset (\textit{regexPRs}) and the merged PRs from prior work (\textit{allPRs}), as described in Section~\ref{sec:analysis:rq2:complexity}. We compare the distributions of each feature across the datasets using a Mann-Whitney-Wilcoxen test of means. For each feature, we present the 5\% percentile,  mean, median, 95\% percentile, and skewness score. The skewness score is calculated according to Pearson's moment coefficient of skewness~\cite{skewness1,skewness2}. 
For example, for the merged pull requests in \textit{allPRs}, the median \textit{num\_commits} is 1 and the skewness is 16.75. 
Although the median number of commits is also 1 in \textit{regexPRs}, the skewness of commits is only 7.97. This means the distribution of \textit{num\_commits} has a shorter tail in \textit{regexPRs}, because of which the 95\% percentile of \textit{num\_commits} in \textit{regexPRs} is smaller than that in \textit{allPRs}. 

As shown in Table~\ref{tab:pr_features}, \textit{regexPRs} has less skewed distributions than  \textit{allPRs} on all features. Therefore, the characteristics of regex-related PRs are less asymmetric than general PRs. 
The Mann-Whitney-Wilcoxon tests between \textit{regexPRs} and \textit{allPRs} show that \textit{regexPRs} take longer to merge (\textit{mergetime\_minutes}) and involve more lines of code (\textit{code\_churn}), and these differences are significant at $\alpha = 0.001$. Our conclusion is that regex-related PRs are different than general~PRs. \\


\noindent \textbf{Summary:} The fixes in regex-related PRs are significantly different from general PRs. Most regex-related PRs take a longer time to get merged and involve more lines of code. 


\subsection{Changes to Regexes in PRs}


 In regex-related PRs, we observed four types of changes: regex addition ($R_{add}$), edit ($R_{edit}$), or removal ($R_{rm}$), or a regex API is modified ($R_{API}$). Table~\ref{tab:FixesOverall} presents the distribution of regex changes over the 356 PR bugs with noted dominant type of regex changes.
Across all root causes and manifestations, the most common change is an edit, as 52.8\% (188/356) of the PRs contain one or more edit. Regexes were added in over twice the number of PRs (119) as they were removed (55).
Regex API changes occurred in 59 (16.6\%) of the PRs. Note that these numbers do not add up to 356 because a PR can have multiple types of changes (e.g., $R_{API}$ and $R_{edit}$);  14.9\% (53/356) of the regex-related PRs involve more than one type of changes. 
Although $R_{edit}$ is the dominant type of regex-related change in our dataset, the number of $R_{edit}$ changes in those pull requests is usually one or two. In contrast, the average number of changes for $R_{API}$ is above seven. 
Next, we examined the fixes applied to each root cause.

\noindent\textbf{Fixes for Regex Root Cause.}
When the regex is the root cause, 78.0\% (170/218) of the PRs contain a regex edit. 
To fix design smells, however, regex removal is more common; 
as 11 of the 17 design smells PRs are related to unnecessary regexes (Table~\ref{tab:hierarchy}), removing the regex is a natural response. 


We note that a regex edit is not always the solution, even when the regex itself is the root cause. 
For example, incorrect regex behavior could be fixed by replacing the regex with an existing parser (See Pattern 4 in Table~\ref{tab:patternsRegex}). 
When incorrect regex behavior relates to the changed input data, the PR can either modify the regex or simply add a regex to the list of regexes (See Pattern 5 \& 6 in Table~\ref{tab:patternsRegex}). 
When the incorrect regex behavior is related to case sensitivity and Unicode characters, adding or modifying the regex flags in the regex API method can also be found together with regex edits (e.g., \href{https://github.com/apache/beam/pull/6092}{apache/beam\#6092}). 


\noindent\textbf{Fixes for Regex API Root Cause.}
Most of the fixes for regex API issues involve changes to the API (78.8\%, 26/33).
Of all the API changes for all root causes (59 PRs, 427 instances), most fix deprecated APIs (71.2\%, 304/427). 
However, multiple changes are sometimes required. For example, the PR~\href{https://github.com/mozilla/treeherder/pull/198}{mozilla/treeherder\#198} handles an incorrect computation and contains an $R_{API}$ and an $R_{edit}$. 
While the fix moves the flag from \mintinline{python}{re.search} to \mintinline{python}{re.compile}, the regular expression \verb!'.+ pgo(?:[ ]|-).+'! is optimized into a different representation \verb!'.+ pgo[ -].+'!, which is a hidden regex representation code smell not mentioned in the PR description. 

\noindent\textbf{Fixes for Other Root Causes.}
The majority (75\%, 173/230) of $R_{add}$ edits come from the \textit{other code} root cause. 
This is fitting as regexes are used in the solution for PRs in this category, but are not the cause of any issues. \\

\noindent \textbf{Summary:}  Suitably, each root cause has a common change type. When regexes are the problem, edits are the most common solution, unless it is a design smell that is resolved through removal. API problems involve API changes, and regexes are often added to solve problems caused by other code. 

\subsection{Recurring Patterns to Fix Regular Expression Bugs}
\label{sec:recurringpattern}
\begin{table}[tpb]
    \caption{Recurring patterns to fix regular expression bugs. Pattern~1-7 are to solve regex issues and Pattern~8-10 are to solve regex API issues. With the exception of Pattern~7 (as noted), each pattern can be applied to each of the languages studied: JavaScript, Python, and Java.}
    \label{tab:patternsRegex}
    \centering
    \begin{tabular}{c| c |r  l}
    \hline
    \textbf{ID} & \textbf{\#PR} & \multicolumn{2}{l}{\textbf{Description with Example Before/After the PR} } \\ \hline
        \multirow{3}{*}{1} & \multirow{3}{*}{17} & \multicolumn{2}{l}{Correctly escaping regex literals} \\ \cline{3-4} 
        & & Before:&\mintinline[fontsize=\small]{Java}{regex="a.png"} \\
        & & After: &\mintinline[fontsize=\small]{Java}{regex="a\.png"} \\ \hline
        
        \multirow{3}{*}{2} & \multirow{3}{*}{17} & \multicolumn{2}{l}{Extend or shrink the character class} \\ \cline{3-4} 
        & & Before: & 
        \begin{minipage}[t]{3.3in}
          \begin{minted}[breaklines, fontsize=\small]{python} 
value_regex = r'[_\w]+'
          \end{minted} 
          \end{minipage}\\
        & & After: & \begin{minipage}[t]{3.3in}
          \begin{minted}[breaklines, fontsize=\small]{python} 
value_regex = r'[_\-\w]+'
          \end{minted} 
          \end{minipage} \\ \hline
        
        \multirow{3}{*}{3} & \multirow{3}{*}{15} & \multicolumn{2}{l}{Replace regex with string methods} \\ 
        & & Before: & \mintinline[fontsize=\small]{python}{if re.match(".*error.*",message):} \\
        & & After: & \mintinline[fontsize=\small]{python}{if "error" in message:} \\ \hline
        
        \multirow{5}{*}{4} & \multirow{5}{*}{11} & \multicolumn{2}{l}{Replace regex with existing parser} \\ \cline{3-4}
        & & Before: & \mintinline[fontsize=\small]{Java}{EMAIL_REGEX_PATTERN.matcher(email).matches();}\\
        & & After: &
          \begin{minipage}[t]{3.3in}
          \begin{minted}[fontsize=\small]{Java} 
import javax.mail.internet.InternetAddress;
InternetAddress emailAddr = new InternetAddress(email);
emailAddr.validate();
          \end{minted} 
          \end{minipage}
          \\ \hline
          
        \multirow{3}{*}{5} & \multirow{3}{*}{10} & \multicolumn{2}{l}{Add or remove a regex alternation} \\ \cline{3-4}
        & & Before: & \mintinline[fontsize=\small]{java}{regex = "win32|windows"} \\
        & & After: & \mintinline[fontsize=\small]{JavaScript}{regex = "wind32|windows|win64"} \\ \hline          
        
         \multirow{7}{*}{6} & \multirow{7}{*}{9} & \multicolumn{2}{l}{Add or remove a regex to the regex list} \\ \cline{3-4}
         & & Before: &   
         \begin{minipage}[t]{3.3in}
         \begin{minted}[fontsize=\small]{python} 
'regexes': [
    re.compile('Ubuntu HW 12.04 x64 .+')
]
          \end{minted} 
          \end{minipage}
         \\
          & & After: &
          \begin{minipage}[t]{3.3in}
          \begin{minted}[fontsize=\small]{python} 
'regexes': [
    re.compile('Ubuntu (ASAN )?HW 12.04 x64 .+'),
    re.compile('^Android 4\.2 x86 Emulator .+'),
]
          \end{minted} 
          \end{minipage}
          \\ \hline
          
        \multirow{3}{*}{7} & \multirow{3}{*}{6} & \multicolumn{2}{l}{Add or remove a regex to the regex representation; Language = \{Python\}} \\ \cline{3-4}
        & & Before: & \begin{minipage}[t]{3.3in}
          \begin{minted}[breaklines, fontsize=\small]{python} 
'pattern': '\d{1,2}/\d{1,2}
          \end{minted} 
          \end{minipage} \\
        & & After: &\begin{minipage}[t]{3.3in}
          \begin{minted}[breaklines, fontsize=\small]{python} 
'pattern': r'\d{1,2}/\d{1,2}'
          \end{minted} 
          \end{minipage}\\ \hline     

        \multirow{4}{*}{8} & \multirow{4}{*}{5} & \multicolumn{2}{l}{Checking null values for regex } \\ \cline{3-4}
        & & Before: &   
         \begin{minipage}[t]{3.3in}
         \begin{minted}[fontsize=\small]{java} 
Matcher matcher = regex.matcher(currentLine);
          \end{minted} 
          \end{minipage}
         \\
        & & After: &
          \begin{minipage}[t]{3.3in}
          \begin{minted}[breaklines, fontsize=\small]{java} 
Matcher matcher = currentLine == null ? null : regex.matcher(currentLine);
          \end{minted} 
          \end{minipage}
          \\ \hline

        \multirow{9}{*}{9} & \multirow{9}{*}{4} & \multicolumn{2}{l}{Regex static compilation} \\ \cline{3-4}
        & & Before: &   
         \begin{minipage}[t]{3.3in}
         \begin{minted}[fontsize=\small]{java} 
String BLACKLIST = "...";
boolean method(String name) {            
        return !(name.matches(BLACKLIST));
}
          \end{minted} 
          \end{minipage}
         \\
        & & After: &
          \begin{minipage}[t]{3.3in}
          \begin{minted}[breaklines, fontsize=\small]{java} 
Pattern BLACKLIST = Pattern.compile("...");
boolean methodE(String name) {            
        return !(BLACKLIST.matcher(name).matches());      
}
          \end{minted} 
          \end{minipage}
          \\ \hline

        \multirow{6}{*}{10} & \multirow{6}{*}{4} & \multicolumn{2}{l}{Conditional checking before regex execution} \\ \cline{3-4}
        & & Before:&   
         \begin{minipage}[t]{3.3in}
         \begin{minted}[breaklines, fontsize=\small]{java} 
Matcher m=Pattern.compile(regex).matcher(currentLine);
          \end{minted} 
          \end{minipage}
         \\
        & & After: &
          \begin{minipage}[t]{3.3in}
          \begin{minted}[breaklines, fontsize=\small]{java} 
if currentLine.contains("error"){
    Matcher m = Pattern.compile(regex).matcher(currentLine);
}
          \end{minted} 
          \end{minipage}
          \\ \hline
    \end{tabular}
\end{table}


Table~\ref{tab:patternsRegex} presents the ten recurring fix patterns we identified from the regex-related pull requests. 
Patterns 1-7 fix regex issues and patterns 8-10  fix regex API issues. 
The column \textit{\#PR} shows the number of pull requests that exhibit the pattern. However, this is not an indication of pattern frequency because a fix pattern can (and does) appear multiple times in the same PR. 
Pattern~7 is language-specific, but the rest are general enough to apply to the three languages: Python, JavaScript, and Java. 

\noindent\textbf{Escaping Issues (Patterns 1 \& 7).}
Pattern 1 fixes incorrect regex behavior and compile errors that result from improper escaping, which we saw in Java, JavaScript, and Python. 
The domain knowledge required in Pattern 1 is to distinguish a regex meta-character from string escape character (e.g., \verb!\b! can be a backspace or a regex word boundary) and from plain text (e.g., `\verb!(!' can be a common left parenthesis or the starting anchor of a regex capturing group).
Pattern 7 is specific to Python and can be used to distinguish regex meta-character escaping (e.g., \verb!\.!) from string character escaping (e.g.,\verb!\n!). 

\noindent\textbf{Regex Scope Issues (Patterns 2, 5 \& 6).}
Pattern 2 adds characters to a character class. 
Pattern 5 and Pattern 6 apply when additional alternatives are needed. When the strings within the regex are expressed in separate regular expressions, they can be combined in a single regex using an OR operator \verb!|!  or grouped into a set of regexes. 

\noindent\textbf{Removing Regexes (Patterns 3 \& 4).}
Pattern 3 replaces the regex using string API functions while Pattern 4 replaces the regex solution with APIs provided in third-party libraries. 
The differences between Pattern 3 and Pattern 4 lie in the matching strings. 
Pattern~4 is used when the matching string has its own syntax grammar (e.g., email address, IP address, URL) and its dedicated parser. 
Pattern~3 is used when the use of string libraries is simpler or easier to understand than the regex implementation, but further research is needed to identify situations when a regex is better and when a string implementation is better. 

\noindent\textbf{Exception Handling (Pattern 8).}
Pattern 8 prevents null values from getting into or out of regex API methods. 
Another fix pattern for regex exception handling uses try-catch code blocks, but this can often be addressed by using smart editors to suggest exceptions to catch, so we omit it from the table. 

\noindent\textbf{Unnecessary Computation (Patterns 9 \& 10).}
Pattern 9 avoids repeated regex compilation by pre-compiling regex objects and making the pre-compiled objects sharable among various functions. 
Pattern 10 reduces the execution frequency of regex methods by conditionally checking the input strings prior to the regex matching. 

\noindent\textbf{Other Patterns.}
Other common patterns include transforming a regex character class into a regex shortcut, adding or removing regular expression anchors, changing regex API, splitting regular expressions apart or merging regular expressions together, 
or switching from capturing groups to non-capturing groups. 
More patterns could be observed, but those presented in Table~\ref{tab:patternsRegex} represent common ones that are candidates for automation based on our careful exploration of the data. \\

\noindent \textbf{Summary:} 
For a regex issue, there are often multiple fix patterns that can help, such as replacing a regex with string library operations or replacing it with external library calls. These patterns provide a first step toward understanding common regex-related code changes, which could enable automated program repair or other automated regex support. 

\section{RQ3 Results: Test Code Characteristics in Regex-related PRs}
\label{sec:rq3}
\begin{table}[!tpb]
\centering
\caption{The number of pull requests w/o different test code changes.}
\label{tab:testchange}
\begin{tabular}{|c|c|c|c|c|}
\hline
\multirow{2}{*}{\begin{tabular}[c]{@{}c@{}}Total\\ PRs\end{tabular}} & \multirow{2}{*}{PRs without test code change} & \multicolumn{3}{c|}{PRs with test code change} \\ \cline{3-5} 
                                                                     &                                               & $T_{add}$          & $T_{edit}$      & $T_{rm}$       \\ \hline
\multirow{2}{*}{356}                                                 & \multirow{2}{*}{182 (51.12 \%)}               & 131 (75.29 \%) & 59 (33.91 \%) & 19 (10.92 \%) \\ \cline{3-5} 
                                                                     &                                               & \multicolumn{3}{c|}{174 (48.88 \%)}            \\ \hline
\end{tabular}
\end{table}
In this section, we present the overview of test code changes in the 356 pull requests. 

\subsection{Overview}
Table~\ref{tab:testchange} shows the overview of test code changes among the 356 regex-related PRs. 
In total, 182 (51.12\%) PRs do not include test code changes while the other 174 PRs (48.88\%) contain changes in test code. 
In 171 PRs with test code changes, we find 984 impacted test cases, including 625 (63.52\%) test cases added, 293 (29.78\%) test cases edited, and 66 (6.71\%) test cases removed.  In the remaining three PRs, there were test code changes detected but it was not clear how many test cases were impacted. 

For the purposes of comparison, we note the number of regex-related PRs with test code changes, 48.88\%, is slightly higher than the reported 33\% in another pull request study~\cite{gousios2014exploratory}. 
However, the differences in context matter as the prior work was conducted with GitHub projects sampled between 2012 and 2013, at a time when PRs were newer to the development process. Differences in the types of contributions and the sizes of the changes can also influence developers' demand for test cases~\cite{pham2013creating}. 

The dominant type of test case change is test code addition ($T_{add}$) as over 75\% of PRs with test code changes contain added test cases (Table~\ref{tab:testchange}). On average, 4.77 test cases are added, 4.97 test cases are edited, and 3.47 test cases are deleted.
The comparison of our dataset with other studies on test code changes~\cite{pinto2012understanding} reveals that the regex-related PRs with test code changes have a higher percentage of test case addition and a lower percentage of test removal. The percentages of $T_{add}$, $T_{edit}$, and $T_{rm}$ reported in~\cite{pinto2012understanding} are 56\%, 29\%, and 15\%. Those percentages in regex-related PRs are 64\%, 30\%, 7\%.

\noindent \textbf{Summary:} Nearly 49\% of regex-related PRs have test code changes. 
The dominant test code change type is test code addition which occurs in 75\% of the regex-related PRs having test code change. 

\begin{table}[!tpb]
\centering
\caption{The distribution of test code changes among the 356 pull requests in each bug root cause and manifestation.}
\label{tab:testRootCause}
\begin{tabular}{|l|l|l|l|l|l|}
\hline
\begin{tabular}[c]{@{}l@{}}Root \\ Cause\end{tabular}                 & Manifestation                                                        & \begin{tabular}[c]{@{}l@{}}PRs with \\ $T_{add}$\end{tabular} & \begin{tabular}[c]{@{}l@{}}PRs with \\ $T_{edit}$\end{tabular} & \begin{tabular}[c]{@{}l@{}}PRs with \\ $T_{rm}$\end{tabular} & \begin{tabular}[c]{@{}l@{}}PRs (\%) with \\ test code changes\end{tabular} \\ \hline
\multirow{3}{*}{Regex}                                                & \begin{tabular}[c]{@{}l@{}}Incorrect \\ Behavior\end{tabular}    & 53 / 165                                                            & 32 / 165                                                             & 10 / 165                                                              & \begin{tabular}[c]{@{}l@{}}79 / 165\\ (47.88\%)\end{tabular}               \\ \cline{2-6} 
                                                                      & \begin{tabular}[c]{@{}l@{}}Compile \\ Error\end{tabular}         & 0 / 8                                                                 & 0 / 8                                                                    & 0 / 8                                                                   & 0 / 8                                                                        \\ \cline{2-6} 
                                                                      & \begin{tabular}[c]{@{}l@{}}Bad \\ Smells\end{tabular}            & 10 / 45                                                             & 7 / 45                                                               & 2 / 45                                                                & \begin{tabular}[c]{@{}l@{}}16 / 45\\ (35.56\%)\end{tabular}                \\ \hline
\multirow{2}{*}{\begin{tabular}[c]{@{}l@{}}Regex\\ API\end{tabular}}  & \begin{tabular}[c]{@{}l@{}}Incorrect \\ Computation\end{tabular} & 2 / 6                                                               & 2 / 6                                                                & 0                                                                     & \begin{tabular}[c]{@{}l@{}}4 / 6\\ (66.67\%)\end{tabular}                  \\ \cline{2-6} 
                                                                      & \begin{tabular}[c]{@{}l@{}}Bad\\ Smells\end{tabular}             & 6 / 27                                                              & 1 / 27                                                               & 1 / 27                                                                & \begin{tabular}[c]{@{}l@{}}7 / 27\\ (25.93\%)\end{tabular}                 \\ \hline
\multirow{3}{*}{\begin{tabular}[c]{@{}l@{}}Other\\ Code\end{tabular}} & \begin{tabular}[c]{@{}l@{}}New \\ Feature\end{tabular}           & 40 / 59                                                             & 9 / 59                                                               & 2 / 59                                                                & \begin{tabular}[c]{@{}l@{}}44 / 59\\ (74.58\%)\end{tabular}                \\ \cline{2-6} 
                                                                      & Bad Smells                                                       & 11 /19                                                              & 3 / 19                                                               & 3 / 19                                                                & \begin{tabular}[c]{@{}l@{}}13 / 19\\ (68.42\%)\end{tabular}                \\ \cline{2-6} 
                                                                      & Other Failures                                                   & 9 / 27                                                              & 5 / 27                                                               & 1 / 27                                                                & \begin{tabular}[c]{@{}l@{}}11 / 27\\ (40.74\%)\end{tabular}                \\ \hline
\multicolumn{2}{|l|}{Total}                                                                                                              & 131 / 356                                                           & 59 / 356                                                             & 19 / 356                                                              & \begin{tabular}[c]{@{}l@{}}174 / 356\\ (48.88\%)\end{tabular}              \\ \hline
\end{tabular}
\end{table}

\subsection{PRs without Test Code Changes}
\label{sec:notTest}
The analysis on PRs without updates to test code seeks to understand the reason some PRs do not have test code changes. We first look at the relationship between the root cause and the test code change among the 356 PRs. 

We observe that PRs associated with certain types of regex-related bugs include few test code changes. 
As shown in Table~\ref{tab:testRootCause}, PRs for regex {\em compile errors} do not involve any test code changes. Since regex compile errors will break all test cases involving the regular expression, often additional tests are not needed. It can also be the case that testing costs outweigh the benefit. For example, the regex edit in PR~\href{https://github.com/apache/nutch/pull/234}{apache/nutch\#234} only impacts one line of code. However, the bug only occurs on Windows-based systems, which would require  a specific test environment to be configured.

We also notice that PRs for solving performance-related bugs often do not have test code changes; these fall under the \emph{Bad Smells} category with \emph{Regex} as the root cause. Seven of the ten PRs that manifested as {\em performance issues} contain no test code changes. 
The other three PRs do contain test code changes; they either add test cases
to test for regex hang bugs or  
adapt existing test cases to reflect the regex changes in the source code. None of the three involve test code changes to measure performance.

PRs for regex code smell {\em regex representation} also do not often contain test code changes as 80\% (8/10) of such PRs do not have test changes. For the other two PRs having test changes, test cases are added because there did not exist tests for the changed regex prior to the PR; test cases are edited as new test scenarios are added to test the same functionality.

Besides the regex-related bug types, the location of the regular expression is another reason for PRs not having test changes. 
The regex change may not be considered as necessary to test if it is irrelevant to the software logic. 
For example, regexes being changed in PR~\href{https://github.com/google/openhtf/pull/347}{google/openhtf\#347} are used in the ``pylint" tool and not in the source code. 
In PR~\href{https://github.com/mozilla/amo-validator/pull/520}{mozilla/amo-validator\#520}, the modified regexes interweave with only the test code to assist running test cases.

Not having test code changes does not mean the test code does not exist. Therefore, test code changes may not be necessary when existing test cases are sufficient to describe the expected behaviors, especially when the regex-related change does not impact its matching behavior. 
For example, the bug described in PR~\href{https://github.com/apache/cordova-ios/pull/376}{apache/cordova-ios\#376} is revealed by the test code. 
Also, for the regex API change of converting regex compilation to static (Patterns 9 in section~\ref{sec:recurringpattern}), running existing test code is enough to make sure it does not impact code functionality but only code execution time.\\

\noindent \textbf{Summary:} PRs of certain types of regex-related bugs (compile errors, performance issues, regex representation) rarely have test code changes. The location of the regular expression and the test code prior to the PR also play a role for developers not to make test code changes.

\begin{table}[tpb]
\centering
\caption{The distribution of test case changes among the 174 PRs with test code changes.}
\label{tab:testchangeDistri}
\begin{tabular}{|l|rrr|r|}
\hline
{} & $T_{add}$ & $T_{edit}$ & $T_{rm}$ & Combined\\
\hline
count &  131 &   59 &  19 & 171 \\ \hline
mean  &    4.77 &    4.97 &   3.47 & 5.75\\ \hline
min   &    1.00 &    1.00 &   1.00  & 1.00\\ \hline
25\%   &    1.00 &    1.00 &   1.00 & 1.00\\ \hline
50\%   &    2.00 &    2.00 &   2.00 & 2.00\\ \hline
75\%   &    5.00 &    3.00 &   3.50 & 5.00\\ \hline
max   &   50.00 &  166.00 &  13.00 & 167.00\\  
\hline
\end{tabular}
\end{table}

\subsection{PRs with Test Code Changes}
To understand the test case changes in the regex-related PRs, we first look at the number of changed test cases, then check its relationship with the regex-related bugs and with the regex changes.

Table~\ref{tab:testchangeDistri} presents the distribution of the number of changed test cases in the 174 PRs with the test code changes. The columns $T_{add}$, $T_{edit}$, $T_{rm}$ shows the number of added, edited, and deleted test cases in the PR, separately. The column {\em combined} shows the total number of the changed test cases regardless of the test case change type. Note that we have 174 PRs that have test code changes but 3 are excluded since we cannot determine their test code changes. 

While 75\% of PRs have the test case changes less or equal to 5, we notice that some PRs have significant test case changes. We examined the six PRs which have more than 20 test case changes. Among them, three PRs modify the list of test inputs that feeds into the test cases, two PRs have at least two completely new test code files, and one PR modifies all test input files of a certain type.

Referring to Table~\ref{tab:testRootCause}, {\em new feature} caused by {\em other code} has the highest percentage of PRs with test code changes (44/59; 74.58\%) and the highest percentage of added test cases (40/44; 90.91\%), followed by PRs for {\em bad smells} of {\em other code} (i.e., 74.58\% (13/19) and 84.62\% (11/13).  
This is in-line with a prior study on test code evolution~\cite{pinto2012understanding} that shows 69\% of the added tests are added to validate newly added code. 
For these two types of PRs, it is highly likely that new code is added into the program since regexes are used for either implementing new features or code refactoring. 
Therefore, the majority of the tests for these two types of PRs are most likely added to validate the new code as well. 
Actually, tests are even requested by PR reviewers before merging it into the code repository (e.g., PR ~\href{https://github.com/apache/beam/pull/5528}{apache/beam\#5528}).

Although $T_{add}$ is the dominant type of test code changes, PRs caused by {\em regex} stand out as they consist of a more balanced distribution of test code changes.
The percentage of $T_{add}$ (66.32\%; 63/95) is smaller than the average 75.29\%; the percentage of $T_{edit}$ (41.05\%; 39/95) is higher than the average 33.91\%; the percentage of $T_{rm}$ (12.63\%, 12/95) is also higher than the average 10.92\%. 
One possible explanation for PRs of {\em incorrect behavior} is that the non-matching behaviors of regexes are not often tested~\cite{peipei2018}. Therefore, when an invalid string is accepted, developers have the option to delete the invalid string or to edit the string into a valid one. 
However, we have not found a plausible explanation for bugs causing regex {\em bad smells} due to the small number of pull requests (16) and the diversity of this bug type. \\

\noindent \textbf{Summary:} 75\% of the PRs having test code changes contain no more than five changed test cases. Although test case addition is the dominant test change type across PRs with different root causes, not all test cases are added to validate bug fixes. Depending on the type of the regex-related bug, test code changes are commonly used to validate new functionality, code refactoring, and bug fixes. 

\section{Discussion and Future Work}
\label{sec:discuss}
We began this work to gain a better understanding of the issues developers face when working with regular expressions, and the lens we chose is the pull request. Here, we discuss our high-level observations, implications, and future work possibilities. 
Based on our analysis of the data, the following observations stand out: 



\noindent\textbf{Differences across programming languages.}
\label{sec:lang}
Prior work shows that the regular expression representations have significant differences across programming languages~\cite{davistesting} and porting regular expressions causes semantic and performance differences~\cite{davis2019aren}. During our analysis of regex bugs, we saw that some regex bugs are closely related to a particular program language. 
The incorrect computation or incorrect regex behavior caused by stateful methods occurs only in JavaScript. 
The Regex API code smell of \emph{Performance/Security} 
occurs in JavaScript and Python, but not Java (Section~\ref{sec:regexapi:smells}).
The language version also has an impact on regex bugs by changing flags (e.g., \mintinline{python}{re.L} is no longer supported after Python 3), deprecating APIs, and changing performance. 



\noindent\textbf{Regex issues when represented as string literals.}
When a regular expression is represented as a quoted string literal, it can be tricky to get right. 
Regexes use backslashes for shortcuts (e.g., \verb!\d!) and to convert meta-characters to plain characters (e.g., \verb!a\.png!). However,  backslashes themselves need to be escaped to make a valid string sequence. The complicated escaping process and the different escaping character support in different languages make regular expression escaping fragile (see Pattern~1 in Table~\ref{tab:patternsRegex}). 


\noindent\textbf{To regex or not to regex.} 
Our study found 15 PRs of replacing regex with string operations and 9 PRs of replacing string operations with regular expressions. 
When other code is the root cause of the issue, regexes are added in 82.9\% (87/105) of the PRs. 
The problem of when regular expressions should and should not be used ~\cite{regexabuse,regexabuse2,stackexchange} is also discussed in the PRs. One PR discussion sets a boundary for when regexes should be used: ``\textit{If the data and the comparison only require you to test for equality, then I'd try to use an Array. If whatever I'm testing can't use equality then I'd use a RegExp.}" (\href{https://github.com/mozilla/fxa-auth-server/issues/1740}{mozilla/fxa-auth-server\#1743}). 
This problem is regarded to be one of the difficulties of regex programming~\cite{michael2019regexes}. Further research efforts are needed to better understand when to use and when not to use regexes.

\noindent\textbf{Regex usage context matters.}
In this paper, we found that regex errors go beyond just composing the intended regex. 
The issues we observed also include incorrect usage of regular expression APIs (Section~\ref{sec:api}), improper exception handling (Section~\ref{sec:regexapi:smells}), and unreadable or inefficient regexes (Section~\ref{sec:regexapi:smells}). 
Thus, it is important to consider regexes in their context when proposing solutions to support developers. Online tools, which developers report to use for regex composition and testing~\cite{chapman2016exploring}, cannot determine if a regex is compiled too often (Pattern~9, Table~\ref{tab:patternsRegex}), if a string library would be more appropriate (Pattern~3, Table~\ref{tab:patternsRegex}), or if a meta-character should be escaped (Pattern~1, Table~\ref{tab:patternsRegex}). While helpful for understanding matching behavior, developers could benefit from tool support within the IDE that can consider the context. 


\noindent\textbf{Regex performance is about more than regex complexity.}
Prior work on regexes and ReDoS~\cite{davis2018impact} focuses on the complexity of executing a regular expression. In the PRs we studied, developers demonstrated an interest in optimizing regex execution by refactoring the surrounding code (e.g., adding conditional or null checking, Patterns 8 \& 10, Table~\ref{tab:patternsRegex}) or by fine tuning the features in the regular expression representation such as changing capturing groups to non-capturing groups (e.g., \href{https://github.com/apache/nutch/pull/432}{apache/nutch\#432}). Automated performance support would help developers identify these inefficiencies sooner. 

\noindent\textbf{Testing regexes is hard and testing regex performance is harder.}
Prior work on regex testing~\cite{peipei2018} shows that only 17\% regular expressions are tested. 
The PRs reveal that test code is not typically committed with regex changes; only 48.88\% of regex-related PRs include test code changes. The percentage of PRs with test changes is much lower than that reported in other bug benchmarks~\cite{madeiral2019bears,gyimesi2019bugsjs}~\footnote{The former reports 94\% of found bug-fixing patches contain test cases and the latter reports only four patches with no tests changed}. While real-world regex bug benchmarks for regex testing and repairing techniques would be beneficial, constructing such benchmarks could be more challenging given the low frequency of regex-related tests~\cite{peipei2018} and the low frequency of bug fixes containing changed tests.


There are very few test case changes regarding the performance impact from regex-related changes. For PRs causing code smells related to performance, the optimization has either no demonstration, is made based on other resources, or is demonstrated by the program execution time prior to and after the PR. 
For example, PR~\href{https://github.com/apache/cxf/pull/479}{apache/cxf\#479} makes the regex changes based on a StackOverflow question and provides that as evidence in the bug description. 
A specialized testing benchmark for regular expressions could help developers make better regex usage optimization.

Note that we do observe some PRs include performance information. For example, PR~\href{https://github.com/apache/druid/pull/3642}{apache/druid\#3642} containing measurements of performance using JMH benchmarks~\footnote{\url{https://openjdk.java.net/projects/code-tools/jmh/}}. 
PR~\href{https://github.com/apache/cxf/pull/479}{apache/cxf\#479} also contains a benchmark to compare the execution time of two Java methods (i.e., \mintinline{java}{String.split} and \mintinline{java}{Pattern.split}). However, the location is the pull request description instead of the test code. 
PR~\href{https://github.com/google/grr/pull/131}{google/grr\#131} and PR~\href{https://github.com/mozilla/treeherder/pull/60}{mozilla/treeherder\#60} report the performance differences before and after the regex modification but without test code changes. 

Through this exploration of test code, we reflect that the regex testing statistics from our prior work~\cite{peipei2018} may be artificially low due to feasibility issues. Not all regexes can be tested in context. 
For example, the regular expressions written in configuration files (e.g., \href{https://github.com/mozilla/amo-validator/pull/520}{mozilla/amo-validator\#520}) or the regular expressions causing compile errors (e.g., \href{https://github.com/apache/nutch/pull/234}{apache/nutch\#234}). 
In that case, it is important to ensure the regexes are not malicious and do not cause significant system slowdown.


\noindent\textbf{Summary:}
Each of these observations opens the door for further research. Our sample of PRs was not large, but the analysis was in-depth. Opportunities for further, automated exploration and further, automated support have been identified. 

 \section{Threats to validity}
\label{sec:threats}
\noindent\textbf{Internal Validity.} 
We manually labeled the PRs using two authors as raters. To reduce misclassifications, all disagreements were thoroughly discussed with a third author. 


Our analysis considers only the code changes present in merged pull requests. Thus, and changes that proceed or follow the PR but are not linked to the PR were not analyzed.


\noindent\textbf{Construct Validity.}
We analyzed 356 merged PR bugs from 4 organizations, which may not be representative of all regex-related PRs. These PRs are in three languages, which may not generalize. The dataset is from public GitHub repositories, which may not generalize to projects hosted elsewhere or private repositories. However, we did not observe any important differences in PRs between the selected organizations.  Their distributions of root causes and manifestations,  are not statistically different from one another, suggesting (though not proving) generalizability. 

When comparing \textit{regexPRs} and  \textit{allPRs}, we observed that 8 PRs in \textit{regexPRs} are present in \textit{allPRs}. We believe the impact is minimal, as there are over 800x more PRs in the \textit{allPRs} dataset. 

We split six PRs into multiple bugs because the issues were independent. This has a subtle impact on the generalizability of the results to other sets of regex-related PRs. 

Where appropriate, we connected our results to prior work on regular expressions to reduce mono-method bias. 

\noindent\textbf{External Validity.} 
The PRs were sampled on February 1, 2019, and thus reflect the PRs available at a specific date and time. Results may not generalize to PRs sampled from a different period.

We used GitHub’s merge status in selecting PRs, which poses a threat to validity~\cite{kalliamvakou2014promises}. This threat is that additional pull requests may have been merged, and the existence of such pull requests would affect our results if they substantially differ from the ones merged via GitHub. Further study is needed to assess the impact of this threat. 


Among the 16 PR features~\cite{gousios2014dataset}, we only select four of them to evaluate RQ2. The comparison between \textit{regexPRs} and the \textit{allPRs} dataset may not hold on the other features. 

As pointed out by Zaidman, et.al~\cite{zaidman2008mining}, in open-source software systems, test code and production code can evolve synchronously following the test-driven development instructions. However, they may also have periods of pure testing and pure development. Since we only examined the test code before the pull requests and the test code changes in the pull request, our observation and conclusion may not hold if the test code for the regex changes are made in a separate commit or pull request posterior to the PR making regex changes.

\section{Related Work}
\label{sec:related}

This work is mostly related to research on regular expressions in software engineering. 
The methodology is most related to research on software bugs and classification. 

\noindent\textbf{Regular Expressions in SE.}
Empirical research on regular expressions in software engineering is emerging (e.g.,~\cite{chapman2016exploring,chapman2017exploring,davis2018impact,peipei2018,peipei2019saner,davis2019aren,davistesting,park2019softregex,zhong2018semregex,chen2020multi}).
Previous research focuses on regex feature usage in one language~\cite{chapman2016exploring} and later on comparing regex characteristics across languages~\cite{davistesting,davis2019aren}, with a specific focus on portability issues~\cite{davis2019aren}.
Previous research also explores regex characteristics (e.g., size, features) at a moment in time~\cite{chapman2016exploring,davis2019aren}, or, more similar to this work, on changes to the characteristics over time through the lens of commit history~\cite{peipei2019saner}.
Another dimension is context: some regular expression studies extract regexes from their context for analysis (e.g., \cite{chapman2017exploring,peipei2019saner}), but others consider the execution environment (e.g., to measure test coverage~\cite{peipei2018} or identify actual ReDoS issues~\cite{davis2018impact}). 
Prior work has also shown researchers' interest in regular expression synthesis with example-based~\cite{bartoli2016inference,ye2019sketch}, NL-based~\cite{zhong2018semregex,park2019softregex,locascio2016neural}, and multi-modal~\cite{chen2020multi} techniques. 
In this work, we analyzed regexes in multiple languages using the context from PRs, which is not available through commit history alone, in addition to properties of the regex itself. 

Regex comprehension has been studied using controlled experiments~\cite{chapman2017exploring} and composition strategies have been studied using observational studies of developers~\cite{bai2019exploring}. This work is complementary to work in regex comprehension, as regex representation code smells were found in this work. These are the byproduct, in part, of regex readability issues  (Section~\ref{sec:regex:smell}).

Complementary to our efforts here, prior work identifies the presence of ReDoS vulnerabilities in thousands of JavaScript and Python modules~\cite{davis2018impact,staicu2018freezing,davisusing}. While the prior work~\cite{davis2018impact} took a deep dive into a particular type of vulnerability, this work looks more broadly at issues resulting from regular expressions (including ReDoS issues, which were also present in two PRs in our dataset, Section~\ref{sec:regex:smell}).

Prior work has surveyed developers to identify pain points associated with regular expression usage~\cite{chapman2016exploring,michael2019regexes}. Rather than surveying developers, this work explored the discussions in regex-related PRs. Pain points were revealed indirectly through the fix patterns (e.g.,~issues with escaping literals are common and likely a pain point), and bug characteristics. 

\noindent\textbf{Software Bugs and Classification.}
GitHub has become a popular hosting site for organizations large and small to make their projects available to their teams and the public. 
Pull requests are created when a developer wants their changes to be integrated into a project; sometimes these are linked to a GitHub issue or another bug reporting software. Pull requests are reviewed and discussed before being merged.

The lens through which researchers study bugs is typically a bug report~\cite{tan2014bug,di2017comprehensive,zhong2015empirical,ma2017developers,zhang2018empirical}.   GitHub pull requests~\cite{gousios2014dataset,majumder2019software,gousios2014exploratory,roy2014insight} provide a different lens as they contain a proposed (or actual, in the case of a merged PR) change. 

Similar research to ours is bug classification~\cite{maalej2015bug,herzig2013s,ohira2016case}. 
Some research  targets emerging applications, such as TensorFlow bugs~\cite{zhang2018empirical} and Blockchain bugs~\cite{wan2017bug}, while others target distributed systems such as node change bugs~\cite{lu2019understanding} and concurrency bugs~\cite{lu2008learning}. More specific bug types include bugs in exception-related code~\cite{coelho2015unveiling}, bugs in patches~\cite{yin2011fixes}, bugs in test code~\cite{vahabzadeh2015empirical}, numerical bugs~\cite{di2017comprehensive}, string-related bugs~\cite{eghbalino}, performance bugs~\cite{selakovic2016performance}, and cross-project correlated bugs~\cite{ma2017developers}. 
Our study joins this list with its focus on regex-related bugs. 
It is interesting to see that 37\% of the string-related bugs~\cite{eghbalino} are caused by regular expressions with six recurring bug patterns. Our study of regex-related bugs and the string-related bug study are complementary to each other.

Bugs are often categorized in terms of root causes and manifestations~\cite{lu2008learning,di2017comprehensive,zhang2018empirical,selakovic2016performance,yin2011fixes}, bug patterns~\cite{lu2008learning,10.1145/3368089.3409760}, and fix strategies~\cite{lu2008learning,ma2017developers,selakovic2016performance}. 
Tan, et al.~\cite{tan2014bug} conduct a temporal analysis to study the trend of different types of bugs with software evolution. 
Zhong and Su~\cite{zhong2015empirical} evaluate the differences between bug fixes by programmers and the fixes by automatic program repair. 
Selakovic and Pradel~\cite{selakovic2016performance} measure the complexity of optimization code changes.
Wan et al.~\cite{wan2017bug} evaluate the relationship between bug type and bug fixing time. 
We adopt the approach of using root causes and manifestations to describe regex-related bugs and the approach of using fix strategies to describe bug resolution. 




In addition to bug studies, there is also lots of research focused on code refactoring to categorize or detect code smells and design smells in source code~\cite{palomba2013detecting,moha2009decor,sharma2016does} and to understand the mutual impact between those bad smells and the software development process~\cite{khomh2009exploratory,tufano2015and}. As many of the PRs were addressing code smells, our work is related to this literature as well.

\noindent\textbf{Test Code Studies.}
Two popular test code study areas are test code evolution and testing behavior. 
The test code evolution can be on either macro-level or the micro-level categorizing why and how the test code evolves. 
Zaidman, et al.~\cite{zaidman2008mining} and Marsavina, et al.~\cite{marsavina2014studying} study the testing strategies and the co-evolution patterns between production code and test code evolution.
Pinto, et al.~\cite{pinto2012understanding} calculate the distribution of added, removed, and modified test cases and investigate the reasons for different test case changes.
While our study of test code is also micro-level and focus on test code addition, deletion, and modification, we focus only on regex-related code changes. 

Testing behavior research has many facets. 
Gousios, et al.~\cite{gousios2014exploratory} report that 33\% of pull requests include test code modifications in 291 selected Ruby, Python, Java and Scala projects. It also finds that the inclusion of test code does not affect the time or the decision to merge a pull request. Kochhar, et al.~\cite{kochhar2013adoption} study the distribution of test cases across 50,000 GitHub projects. 
Pham, et al.~\cite{pham2013creating} is a study of testing behaviors on GitHub and points out that developers' demand for test code in pull requests is influenced by the size of code changes, the types of contributions, and the estimated effort.
In our work, we utilize the test case detection conventions listed in prior work~\cite{gousios2014dataset} and manually find the test case changes. 
Test code changes are also an important factor in constructing real-world bug benchmarks. Since the availability of benchmarks facilitates software testing, debugging, and automated repairing techniques, changed test cases are identified to guarantee the reproducibility of bugs~\cite{widyasari2020bugsinpy,just2014defects4j,gyimesi2019bugsjs,madeiral2019bears,saha2018bugs}.

\section{Conclusion}
\label{sec:conclusion}
The regular expression is one of the primary culprits of string-related bugs. While it is acknowledged that regular expressions are difficult to use correctly and poorly tested, there is little knowledge about what regular expression problems could happen in real-world software code and the consequences of those problems. Prior work exploring why regular expression testing is difficult is also scarce. 

This paper presents a comprehensive study of 356 merged regular expression related pull requests from Apache, Mozilla, Facebook, and Google GitHub repositories where the regular expression problems are studied carefully via bug classification, bug fix, and the test code in the bug fix. 
The results of this study include: 1) a spectrum of regular expression root causes and manifestations with their frequency in real-world software; 
2) ten common patterns of regex bug fixes; 
and 3) quantitative and qualitative analysis of pull request test code changes. 
We demonstrate that regular expression problems are not trivial, as regex-related PRs take more time and more lines of code to fix compared to general pull requests. 
Our study shows that regular expression bugs are not independent of the source code, but are influenced by the software evolution and the code quality. 
The analysis on test code changes indicates that some difficulty in testing regexes lies in the location of the regex and that regex is seldom to be tested alone.
Our results and findings provide an overview of regular expression bugs and motivates future work on techniques and tools to solve practical regular expression problems. 

\begin{acknowledgements}
This material is based upon work supported by the National Science Foundation under Grant No. \href{https://www.nsf.gov/awardsearch/showAward?AWD_ID=1714699}{1714699}.
\end{acknowledgements}

%
%

\bibliographystyle{spmpsci}      
\bibliography{refs}   

\end{document}